\documentclass[twocolumn,fleqn,,usenatbib]{mnras}
\usepackage{hyperref}
\usepackage{bm}
\usepackage{multirow}
\usepackage{gensymb}
\usepackage{enumitem}
\usepackage{tabularx}
\usepackage[flushleft]{threeparttable}
\usepackage[]{color}

\usepackage{float}
\usepackage{breqn}
\usepackage{amsmath}

\usepackage{amsfonts,amsmath,amssymb,esint}
\usepackage{graphicx}
\usepackage{color}
\usepackage{xcolor}
\usepackage{url}

\title[Flux-Averaged Force Multipliers]{Flux-Averaged Force Multipliers}

\author[S. Dyda et al.]{
Sergei Dyda,$^{1}$\thanks{sdyda@ua.edu}
Randall C. Dannen,$^{2,3}$
Shane W. Davis,$^{4,5}$
Daniel Proga$^{2,3}$
and Timothy R. Kallman,$^{6}$
\\
$^{1}$ Department of Physics \& Astronomy, University of Alabama, Gallalee Hall, 514 University Blvd, Tuscaloosa, AL 35401, USA \\
$^{2}$ Department of Physics \& Astronomy, University of Nevada, Las Vegas, 4505 S. Maryland Pkwy, Las Vegas, NV, 89154-4002, USA \\
$^{3}$ Nevada Center for Astrophysics, University of Nevada, Las Vegas, 4505 S. Maryland Pkwy, Las Vegas, NV 89154, USA \\
$^{4}$ Department of Astronomy, University of Virginia, 530 McCormick Rd., Charlottesville, VA 22904, USA \\
$^{5}$ Virginia Institute for Theoretical Astronomy, University of Virginia, Charlottesville, VA 22904, USA \\
$^{6}$ NASA Goddard Space Flight Center, Greenbelt, MD 20771, USA
}

\begin{document}

\label{firstpage}
\pagerange{\pageref{firstpage}--\pageref{lastpage}}

\maketitle

\begin{abstract}
We apply novel developments in photoionization modeling and multi-frequency radiation hydrodynamics to the study of line driven AGN disc winds. We use a flux-averaged force multiplier approach to compute the radiation force due to lines for hydrodynamics simulations using 4 frequency bands - infrared (IR), optical (O), ultraviolet (UV) and X-rays. Though line driving is dominated by the UV, contributions from the O and X-ray bands are non-negligible and can lead to enhancements in the wind both in terms of mass flux and outflow velocity. Crucially, these effects are not captured when using a ``grey'' approach to the radiation modeling in the hydrodynamics, where frequency information is averaged over during the photoionization modeling. These results further strengthen the case for frequency dependent radiation dynamics studies for line driven winds.

\end{abstract}

\section{Introduction}
\label{sec:introduction}

Radiation plays a fundamental role in launching astrophysical outflows from compact objects by either heating the outflowing material or directly transferring momentum to it via radiation pressure \cite{Begelman83,CAK1975}. These processes are at play in various astrophysical systems such as young stellar objects, X-ray binaries, and active galactic nuclei (AGN). AGN are a prime example, as the radiative environment is particularly intense, with systems close to the Eddington limit. The radiation pressure, especially in the ultraviolet (UV), can overcome the gravitational force from the central potential, driving large-scale outflows. These AGN-driven winds can reach velocities of thousands of kilometers per second \citep{Chartas02,2010A&A...521A..57T,Pounds13}, influencing galaxy evolution by regulating star formation and redistributing interstellar material \citep{Silk1998,Fabian2012,Harrison2018}.

Different types of radiation influence astrophysical outflows in distinct ways. X-rays and extreme UV, primarily act by ionizing the surrounding gas. This ionization heats the gas and alters the ionic level populations, affecting the gas opacity. This opacity is dominated by spectral lines in the UV band, and therefore UV photons can efficiently transfer momentum to the gas via radiation pressure, a process known as line driving \citep{Lucy1970, CAK1975, 1994ApJ...427..700A}. UV photons are absorbed by bound electrons in ions, imparting momentum and accelerating the gas outward. This selective absorption is highly sensitive to the ionization state of the gas, creating a delicate balance where increasing the UV and X-ray flux can provide additional momentum for the gas but also suppress any momentum transfer by decreasing the gas opacity \citep{Murray1995, 2003MNRAS.344..233C}. 

The irradiating spectral energy distribution (SED) thus plays a dual role for the gas dynamics. Firstly, it acts as a source of momentum, allowing the gas to accelerate and escape the central object's gravitational potential. Crucially, it plays a second, equally important role, determining the strength of the gas/radiation coupling  and thus the efficiency of this momentum transfer. This occurs on microphysical scales, where the irradiating SED determines the level populations of ionic species in the gas and consequently the opacity of the gas. It is via this gas opacity that photons are scattered, absorbed and re-emitted, dynamically affecting the gas. Given the dual role of the SED, both as a source of momentum available to the gas but also as the driver of the gas/radiation coupling, astrophysical wind models should aim to account for SED effects in as self consistent way as possible \citep{Chakravorty09}.

Ionization effects are typically modeled using a photionization code such as \textsc{XSTAR} \citep{XSTAR2001} or \textsc{Cloudy} \citep{2013RMxAA..49..137F} which simulate the physical conditions of gas exposed to a radiation field. They take inputs such as the shape and intensity of the ionizing spectrum, gas density, and elemental abundances, and then solve for the ionization balance, temperature, and resulting emission or absorption spectra. These codes calculate how radiation ionizes atoms, heats the gas, and produces observable features, helping to model and interpret astrophysical outflows, nebulae, and other ionized environments. Outflows are studied using hydrodynamics codes, which solves the equations of fluid dynamics to simulate the motion of gas under forces like pressure gradients, gravity, and magnetic fields. 

Due to computational constraints, the microphysics of the gas in hydrodynamics simulations must often be simplified either through analytic approximations \citep{Blondin1990,Blondin1994}, precomputed lookup tables spanning the relevant parameter space \citep{Dyda17}, or by calculating it on-the-fly within the hydrodynamic simulation but at a much lower temporal resolution than the main solver \citep{Higginbottom2020}. An alternative emerging strategy leverages machine learning, where models are trained on outputs from photoionization codes to rapidly predict heating and cooling rates \citep{2024arXiv240619446R}. These gas microphysics prescriptions are then implemented in the hydrodynamics simulations.
 
These methods have been applied to the study of AGN line driven disc winds. \cite{PSK2000} used the analytic fits for the photoionization model in \cite{1991ApJ...379..310S} to model the line force. They assumed the disc emitted like a Shakura-Sunyaev disc, and all disc photons contributed to the line force. Photons from a central point source served as a source of ionizing radiation and determined the ionization structure. \cite{PK04} sought to improve on this model, by assuming only UV photons from the disc contributed to the line force. \cite{DDP24} used time-dependent radiation transfer to model the ionizing X-ray source, allowing for absorption, scattering and re-emission. \cite{2025arXiv250400117D} (hereafter DDD25) improved on the treatment of the line force, by using the force multiplier calculations of \cite{Dannen19} (hereafter D19). D19 importantly found that X-ray lines, are present in similar numbers and oscillator strength as UV lines and should not be neglected. In parallel, other numerical studies have explored line-driven AGN outflows under various conditions, reinforcing these findings \citep{2009ApJ...693.1929K, Nomura2016, Nomura2020, Matthews16, Matthews2020}

Monte Carlo radiation transfer (rad-MC) is a powerful technique used to model how radiation interacts with gas in radiation-driven winds by simulating the paths of a large number of photon packets as they scatter, absorb, and re-emit throughout the medium.   This stochastic approach is particularly well suited for handling complex geometries and frequency-dependent opacities, making it ideal for studying outflows where line and continuum processes play a critical role. rad-MC methods have been used to study a variety of outflowing astrophysical systems, notably mass loss rates and winds structures from massive stars, and disc winds from AGN \citep{Matthews2023}.   More recently, these techniques have been adapted to time-dependent studies of active galactic nuclei (AGN), where the radiation field and wind geometry can vary significantly over time \citep{Matthews2025}. rad-MC comes with its own set of challenges, primarily the computational cost. The statistical nature of the method requires a large number of photon packets to propagate through the flow. Further, the separation of scales between the light and sound crossing times makes it challenging to couple the hydrodynamics and radiation transfer in a time-dependent way.  

The state-of-the-art fundamentally suffers from the fact that the SED used in the photoionization studies are inconsistent with the SED in the hydrodynamics modeling. DDD25 used the photoionization tables in D19, which assumed a \emph{global} SED irradiating the flow, notably the spectra for the obscured AGN NGC 5548 \citep{Mehdipour15}. Purely from geometric factors, not even considering absorption, scattering or re-emission effects, the SED irradiating different parts of the flow is very different from this assumed global SED \citep{Smith24}. \cite{2025arXiv250322799D} (hereafter Dyda25) explicitly showed using hydrodynamics simulations that the strength of line driven winds irradiated by blackbody spectra are altered when one accounts for the irradiating SED in a self-consistent way.

We summarize these findings that the \emph{local} irradiating SED can have an important effect on the resulting outflow, and that the SED is highly position dependent for AGN disc winds. This study seeks to overcome the limitations of previous works such as DDD25 by accounting for the effects of position dependent SEDs by using a flux-weighted average for the local force multiplier. This allows one to account for local changes in flux from different frequency bands, but assumes the local SED has no effect on the ionic level populations.

The structure of this paper is as follows. In Sec \ref{sec:theory} we describe our treatment of the line force using the flux-average approximation. We then show in Sec \ref{sec:ForceMultiplierMaps} how this treatment affects the strength of the line force when applied to the geometry of AGN disc winds. We then use these treatments of the line force to perform a series of hydrodynamics simulations, Sec \ref{sec:HydroSims}. We discuss our findings in Sec \ref{sec:discussion} and lay out future directions for studying these systems.

\section{Theory}
\label{sec:theory}
We first describe the general theory of the radiation force due to spectral lines, so called line driving, by deriving some expressions for the force multiplier across many frequency bands. We then describe our photoionization modeling using \textsc{XSTAR} to compute force multipliers for template AGN SEDs. We finally describe an approximation schemes for implementing this photoionization modeling into a hydrodynamics code.

\subsection{Force Multiplier Calculations}
\label{sec:forceMultiplier}

The force per unit mass due to an individual line at frequency $\nu_0$ can be computed as 
\begin{equation}
    f_L = \frac{F_{\nu} \Delta \nu_D}{c} \frac{\kappa_L}{\tau_L} \left( 1 - e^{-\tau_L}\right),
\label{eq:aline}    
\end{equation}
where $\kappa_L$ is the line's opacity, $\tau_L$ the optical depth, $\Delta \nu_D = \nu_0 v_{\rm{th}}/c$ the thermal Doppler width, with $v_{\rm{th}}$ the thermal speed of the ion corresponding the the atomic line in question and $F_{\nu}$ is the specific flux. The line's optical depth in a static atmosphere is
\begin{equation}
\tau_L = \int_r^{\infty} \rho \kappa_L dr,
\end{equation}
where $\rho$ is the gas density, whereas in an accelerating flow is
\begin{equation}
    \tau_L = \rho \kappa_L l_{\rm{Sob}},
\end{equation}
where the Sobolov length $l_{\rm{Sob}} = v_{\rm{th}} / |dv/dl|$ with $dv/dl$ is the velocity gradient of the flow along the path of the radiation. We re-express the Sobolev length in terms of the optical depth parameter
\begin{equation}
    t = \sigma_e \rho l_{\rm{Sob}}. 
\end{equation}
For a single line, the opacity due to stimulated emission, in units of $\rm{cm^2 g^{-1}}$ is
\begin{equation}
    \kappa_L = \frac{\pi e^2}{m_e c} gf \frac{N_L/g_L - N_U/g_U}{\rho \Delta \nu_D},
\end{equation}
where $gf$ is the oscillator strength and $N_L$ and $N_U$ are the occupation numbers of the lower and upper levels respectively. Re-writting the optical depth as
\begin{equation}
    \tau_L = \eta t,
\end{equation}
where $\eta = \kappa_L / \sigma_e$, we can express the total acceleration as a sum over the individual line contributions (\ref{eq:aline}), 
\begin{equation}
    a_{\rm{tot}} = \frac{\sigma_e}{c}\sum_{\rm{lines}} \frac{F_{\nu} \Delta \nu_D}{\sigma_e} \frac{\kappa_L}{\tau_L} \left( 1 - e^{-\tau_L}\right).
\end{equation}
We can break up this sum over all lines into two summations, one over a set of frequency bands, and a sum over the lines in each of those respective bands
\begin{equation}
    a_{\rm{tot}} = \frac{\sigma_e}{c}\sum_{\nu} \sum_{\mathrm{lines},\nu} \frac{F_{\nu} \Delta \nu_D}{\sigma_e} \frac{\kappa_L}{\tau_L} \left( 1 - e^{-\tau_L}\right).
\end{equation}
Finally, we re-write our expression as
\begin{equation}
    a_{\rm{tot}} = \frac{\sigma_e}{c}\sum_{\nu} F_{\nu} \left( \frac{F}{F_{\nu}}\right) \sum_{\mathrm{lines},\nu} \frac{F_{\nu}}{F} \frac{\Delta \nu_D}{\sigma_e} \frac{\kappa_L}{\tau_L} \left( 1 - e^{-\tau_L}\right).
    \label{eq:a_nu}
\end{equation}
We can simplify the presentation of equation (\ref{eq:a_nu}) by expressing it as
\begin{equation}
    a_{\rm{tot}} = \frac{\sigma_e}{c}\sum_{\nu} F_{\nu} \frac{1}{f_{\nu}} M_{\nu}(t,\eta),
    \label{eq:a_nu_final}
\end{equation}
where 
\begin{equation}
    f_{\nu} = \left( \frac{F_{\nu}}{F}\right),
\end{equation}
is the fraction of the flux in band $\nu$ and 
\begin{equation}
    M_{\nu}(t,\eta) = \sum_{\mathrm{lines},\nu} \frac{F_{\nu}}{F} \frac{\Delta \nu_D}{\sigma_e} \frac{\kappa_L}{\tau_L} \left( 1 - e^{-\tau_L}\right).
    \label{eq:Mnu}
\end{equation}
is the force multiplier for band $\nu$. We note that summing over the $M_{\nu}(t,\eta)$ across all bands, we recover the usual expression for the force multiplier 
\begin{equation}
    M(t,\eta) = \sum_{\nu} M_{\nu}(t,\eta) = \sum_{\rm{lines}} \frac{\Delta \nu_D F_{\nu}}{F} \frac{1}{t} \left(1 - e^{- \eta t}\right).
    \label{eq:Mtotal}
\end{equation}

So far, this derivation has been exact in the sense that it only makes use of the assumptions of the Sobolev approximation. However, to utilize equation (\ref{eq:a_nu_final}) in a fully self-consistent way, the same radiation fluxes $F_{\nu}$ should be used throughout the expression. In other words, the fluxes used in the photoionization modeling (to compute $M_{\nu}(t,\xi)$) and in the hydro simulations (to compute the radiation force on the gas) should be the same. Such a calculation is too computationally expensive and necessitates some approximation scheme to make it more computationally tractable.   

Here we propose one such scheme, which we will refer to as the \emph{flux-averaged force multiplier}. With a flux-averaged force multiplier, we assume a fixed SED when computing the frequency dependent force multiplier (via eq. (\ref{eq:Mnu})), but account for local changes in driving flux $F_{\nu}$ when calculating the force due to lines. In other words, only the explicitly appearing $F_{\nu}$ term in equation (\ref{eq:a_nu_final}) is allowed to vary in the hydrodynamics. Physically, this approximation can be understood as the gas/radiation coupling is insensitive to local variations in the irradiating SED and changes in the line force are due solely to variations in driving flux. Consider a variation in flux $F_{\nu} = F_{0,\nu} + \delta F_{\nu}$ and substituting in (\ref{eq:a_nu_final}), 
\begin{equation}
        a_{\rm{tot}} \approx \frac{\sigma_e}{c}\sum_{\nu} F_{\nu} \frac{1}{f_{\nu}} M_{\nu}(t,\eta) \left( 1 + \frac{\delta F_{\nu}}{F_{\nu}} + \frac{\delta F_{\nu}}{M_{\nu}}\frac{\partial M_{\nu}}{\partial F_{\nu}}\right),
\end{equation}
where we neglect second order terms in $\delta F_{\nu}$. The flux-averaged approximation amounts to neglecting the third bracketed term, in favor of the second i.e.
\begin{equation}
    \sum_{\nu} \frac{\delta F_{\nu}}{M_{\nu}}\frac{\partial M_{\nu}}{\partial F_{\nu}} \ll \sum_{\nu} \frac{\delta F_{\nu}}{F_{\nu}},
\end{equation}
and assuming that it is safe to neglect non-linear terms. Physically, this approximation means that changes in flux in the different bands are assumed to change the available momentum in the radiation field that can be transferred to the gas, but the efficiency of this momentum transfer does not. In other words, small radiation flux changes do not alter the force multiplier, only the radiation force on the gas.

\subsection{Photoionization Modeling}
\label{sec:photoionization}

\begin{figure*}
    \centering
    \includegraphics[scale=0.8]{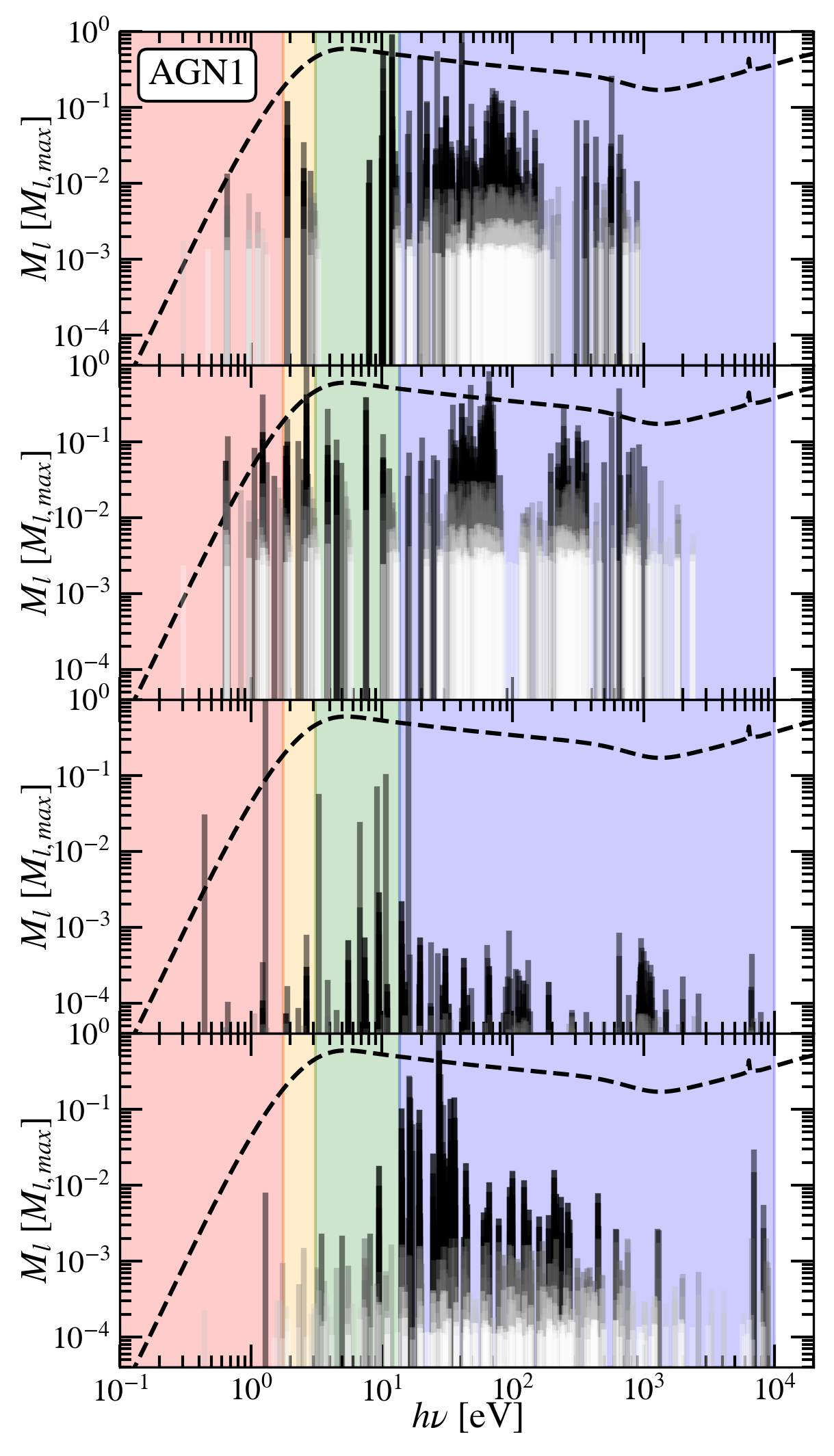}
    \hspace{-11pt}
    \includegraphics[scale=0.8]{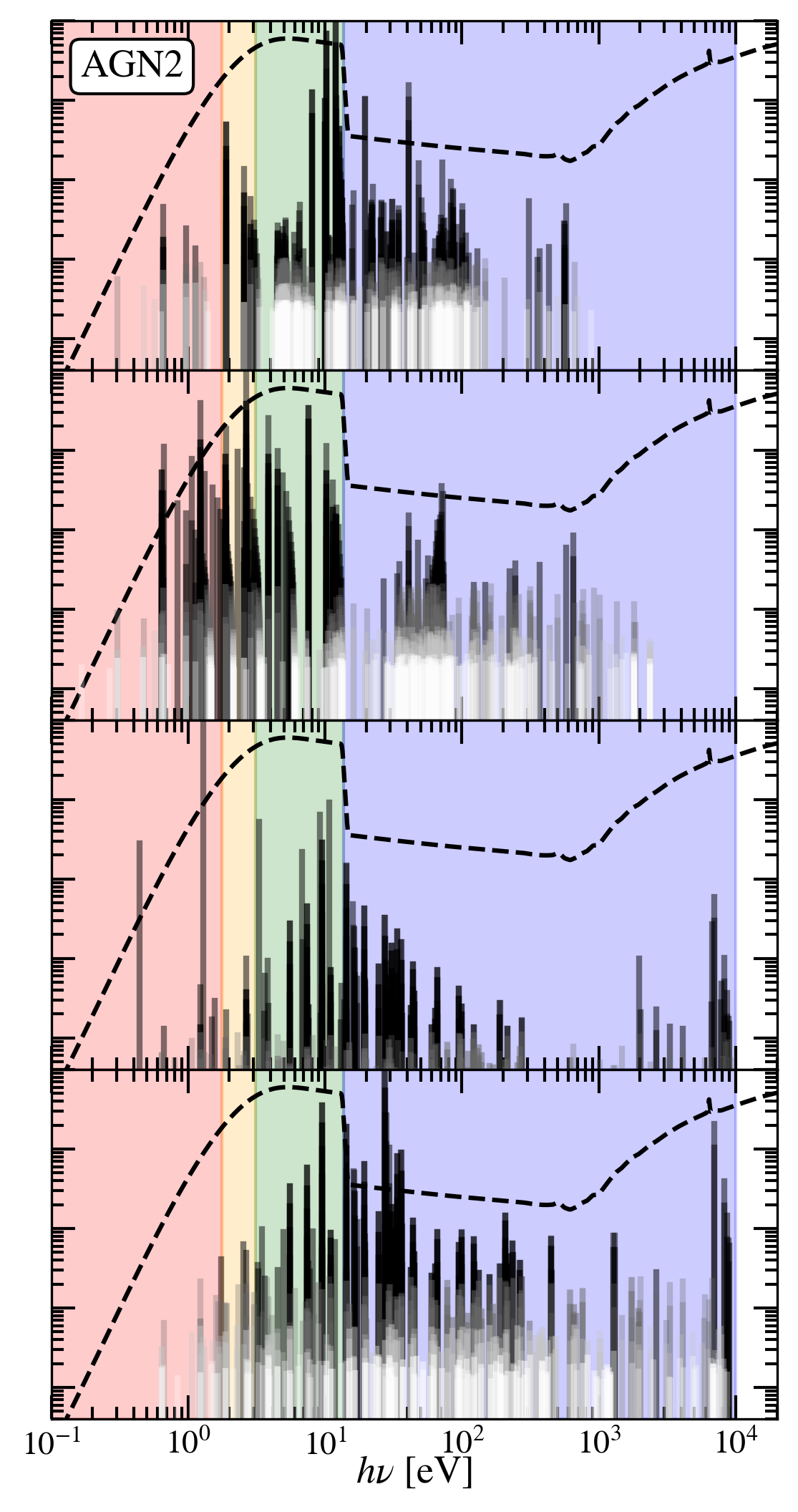}
    \caption{Comparisons of the energy distribution of AGN spectra (black-dashed curves) and single line force multipliers  (grey scale vertical bars) for (in descending order) $\log \xi = 1, 2, 3$ and 4. The vertical bars represent the 1,000 strongest individual line force multipliers – normalized to the maximum force multiplier $M_{\rm{l,max}}$. The maximum force multiplier varies widely at different ionizations, with $\log M_{\rm{l,max}} = 2.51, 1.78, 1.81, 0.68$ for AGN1 and $\log M_{\rm{l,max}} = 2.67, 2.05, 0.25, 2.35$ for AGN2. with increasing ionization parameter. These 1,000 strongest lines have been binned into quartiles, black being the upper quartile, dark grey the second, light grey the third, and white the weakest set of lines. The black dashed line in each panel represents the assumed SED, either AGN1 (left panels) or AGN2 (right panels) that we used for both determining the ionization balance and computing $M$. The colored shading shows the frequency bands listed in Table \ref{tab:bands}.}
    \label{fig:LineDistribution}
\end{figure*}

\begin{figure*}
    \centering
    \includegraphics[scale=0.55]{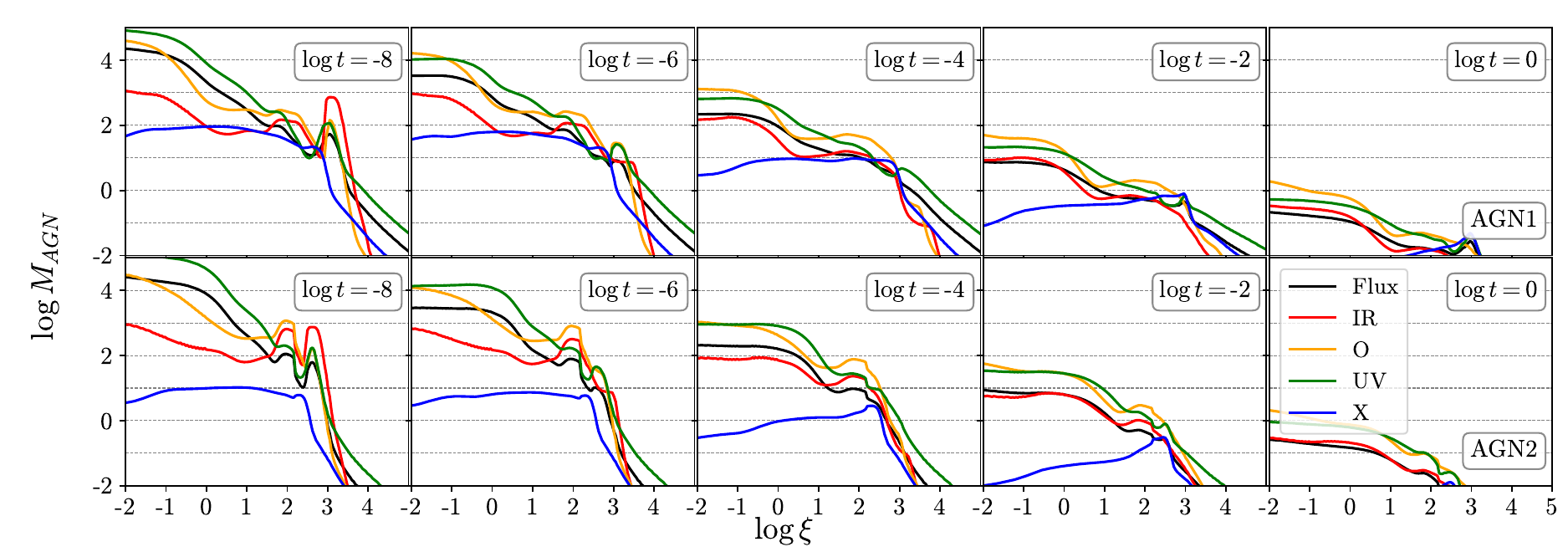}
    \caption{Force multiplier $M_{\nu}$ contributions from the IR (red), optical (orange), UV (green) and X-ray (blue) as well as the flux average (black) as a function of ionizations for a selection of optical depth parameters for an unobscured (AGN1) and obscured (AGN2) SEDs.}
    \label{fig:AGN_4band}
\end{figure*}

\begin{table}
\centering
\begin{tabular}{|c|c|c|c|c|cc|}
\hline
\textbf{Band} & \shortstack{\textbf{$\lambda_{\rm max}$} \\ (Å)} & \shortstack{\textbf{$E_{\rm min}$} \\ (eV)} & \shortstack{\textbf{$\lambda_{\rm min}$} \\ (Å)} & \shortstack{\textbf{$E_{\rm max}$} \\ (eV)} & \multicolumn{2}{c|}{\textbf{$f_i$}} \\
\cline{6-7}
&&&&& AGN1 & AGN2 \\
\hline
IR  & $\infty$ & 0     & 7000  & 1.77   & 0.01 & 0.02 \\
O   & 7000     & 1.77  & 4000  & 3.10   & 0.04 & 0.06 \\
UV  & 4000     & 3.10  & 917   & 13.6   & 0.26 & 0.17 \\
X   & 917      & 13.6  & 1.24  & 10000  & 0.32 & 0.28 \\
\hline
\end{tabular}
\caption{Summary of wavelength and energy ranges for the bands used in this study, with band fractions $f_i$ list for AGN1 and AGN2 SEDs. Any remaining flux is in the hard X-rays, $\lambda < 1.24$ \AA and is assumed to not contribute to the radiation force.}
\label{tab:bands}
\end{table}

We employ the photoionization code \textsc{XSTAR} to compute a grid of photoionization models to implement microphysics into our hydro simulations. For a wide range of optical depth parameter $-8 \leq \log t \leq 1$ and photoionization parameter $-2 \leq \log \xi \leq 5$, we compute the force multiplier (see D19, for a full description of our methods) for a gas irradiated by the AGN spectra in \cite{Mehdipour15}. Notably, D19 used an updated version of \textsc{XSTAR}, including a line list featuring over 2 million lines. We assume solar abundances for atomic species. 

Fig \ref{fig:LineDistribution} shows the energy distribution of single line force multipliers (grey scale vertical bars) for (in descending order) $\log \xi = 1, 2, 3$ and 4 relative to the AGN1 (left panels) or AGN2 (right panels) SEDs (dashed black lines). The vertical bars represent the 1,000 strongest individual line force multipliers normalized $M_{\rm{l,max}}$ for the choice of $\xi$ in that panel i.e. the most significant terms in the sum in equation (\ref{eq:Mtotal}). These 1,000 strongest lines have been binned into quartiles, black being the upper quartile, dark grey the second, light grey the third, and white the weakest set of lines. The colored shading shows the frequency bands listed in Table \ref{tab:bands}.

These plots illustrate the ``line mismatch'' problem \citep{Gayley2000}, where the energies of the strongest lines may not correspond to the peak energies in the irradiating SED. The two AGN SEDs differ in the amount of flux in the soft X-ray. At $\log \xi = 1$, decreasing the soft X-ray flux shifts most of the strongest lines from the soft X-ray to the UV. However, at higher ionizations $\log \xi =4$, this same drop in flux does not significantly affect the lines in the soft X-ray. Thus, a naive study of line strengths may erroneously conclude that line driving is weak (strong), because it computes an ``average'' force multiplier from the convolution of an SED which has weak (strong) flux at the energies of the strongest lines. In a fully self-consistent treatment, the photoioinization modeling would use the local irradiating SED to compute \emph{both} the strength of lines and the radiation flux. Here we use an intermediate approach where we use a global SED to compute the distribution of lines but use a local SED to determine the radiation field providing momentum to the gas.

In order to implement this grid of models in our hydrodynamics scheme, we compiled the data into tables that could be interpolated over as we ran the hydro simulations. We have previously used this method for incorporating heating and cooling (see for example \cite{Dyda17}) and force multipliers (see for example \cite{Dannen20}, \cite{2025arXiv250400117D}). In these previous works, the heating and force multiplier were functions of the ionization state of the gas, temperature or optical depth parameter. Here, the novel aspect is we also include spectral information from the SED and compute the line force contributions $M_{\nu}$ from 4 bands (see equation (\ref{eq:Mnu})), for the infrared (IR), optical (O),  ultraviolet (UV) and X-ray (X) bands ($M_{IR}$, $M_{O}$, $M_{UV}$ and $M_{X}$ respectively, see Sec \ref{sec:forceMultiplier}, eq (\ref{eq:Mnu}) for details). The wavelength/energy ranges for these bands is shown in Table \ref{tab:bands}. 

Fig \ref{fig:AGN_4band} shows the frequency dependent force multipliers $M_{\nu}$ as a function of gas ionization parameters for a representative sample of optical depth parameters for type I (upper panels) and type 2 (lower panels) AGN SEDs. Each color represents a different frequency band and the black line is the flux-averaged force multiplier for each of the assumed SEDs. DDD25 used this flux-averaged SED in the entire simulation domain. From these plots we can anticipate the physical effects which may emerge from this multiband treatment. Directions irradiated by UV photons (and in some parts of parameter space optical photons) will drive gas more strongly than in our prior grey opacity treatment. Positions where certain frequency bands dominate (such as near the disc where the local SED is approximately blackbody and may be peaked around say the optical band) the radiation force will be entirely dominated by a single band.

\subsection{Hydrodynamics \& Radiation Transfer}
\label{sec:hydro}

The equations for single fluid,  multi-frequency radiation hydrodynamics are
\begin{subequations}
\begin{equation}
\frac{\partial \rho}{\partial t} + \nabla \cdot \left( \rho \mathbf{v} \right) = 0,
\end{equation}
\begin{equation}
\frac{\partial (\rho \mathbf{v})}{\partial t} + \nabla \cdot \left(\rho \mathbf{vv} + \sf{P} \right) =  \mathbf{G} + \rho \mathbf{g}_{\rm{grav}},
\label{eq:momentum}
\end{equation}
\begin{equation}
\frac{\partial E}{\partial t} + \nabla \cdot \left( (E + P)\mathbf{v} \right) = cG^{0} + \rho \mathbf{v} \cdot \mathbf{g}_{\rm{grav}},
\label{eq:energy}
\end{equation}
\label{eq:hydro}%
\end{subequations}
where $\rho$ is the fluid density, $\mathbf{v}$ the velocity, $\sf{P}$ a diagonal tensor with components $P$ the gas pressure. The total gas energy is $E = \frac{1}{2} \rho |\mathbf{v}|^2 + \mathcal{E}$ where $\mathcal{E} =  P/(\gamma -1)$ is the internal energy and $\gamma$ the gas constant. The gravitational source is due to a star with
\begin{equation}
    \mathbf{g}_{\rm{grav}} = -\frac{GM}{r^2} \hat{r}, 
\end{equation}
where $M$ is the stellar mass and $G$ the gravitational constant. The temperature is $T = (\gamma -1)\mathcal{E}\mu m_{\rm{p}}/\rho k_{\rm{b}}$ where $\mu = 1$ is the mean molecular weight and other symbols have their standard meaning.  

The radiation source terms $\mathbf{G}$ and $cG^0$ are assumed to receive contributions from the continuum and spectral lines
\begin{subequations}
\begin{equation}
\mathbf{G} = \mathbf{G}_{\rm{cont.}} + \mathbf{G}_{\rm{lines}}, 
\end{equation}
\begin{equation}
G^{0} = G^{0}_{\rm{cont.}} + G^{0}_{\rm{lines}}. 
\end{equation}    
\end{subequations}

The continuum radiation field consists of the X-ray band and are treated by directly solving the time dependent radiation transport equation using the implicit implementation in \textsc{Athena++} \citep{Jiang2022}.  The radiation transfer equation solved is equivalent to
\begin{dmath}
\frac{\partial I}{\partial t} + c \mathbf{n} \cdot \nabla I = c S_{I}, 
\label{eq:dIdt}
\end{dmath}
with the source term
\begin{equation}
\begin{aligned}
    S_{I} = \Gamma^{-3} \rho \Big[ \Big( \kappa_{P} \frac{c a T^4}{4\pi} -  \kappa_{E} J_{0} \Big)
     - \left( \kappa_{s} + \kappa_{F} \right) \left(I_{0} - J_{0} \right) \Big],
     \end{aligned}
    \label{eq:source}
\end{equation}
where $\kappa_{s}$ is the scattering opacity, $\kappa_{F}$ is the absorption contribution to the flux mean opacity, $\kappa_{P}$ the Planck mean and $\kappa_{E}$ the energy mean opacity. $I_{0}$ is the intensity in the comoving frame and 
\begin{equation}
    J_{0} = \frac{1}{4\pi}\int I_{0} \; d\Omega_0,
\end{equation}
is the corresponding angle averaged comoving frame mean intensity.

With the above assumptions the continuum momentum and energy source terms are then
\begin{equation}
    \mathbf{G}_{\rm{cont.}} = \frac{1}{c} \int \mathbf{n} S_{I} d \Omega,
\end{equation}
\begin{equation}
    cG^0_{\rm{cont.}} = c  \int S_{I} d \Omega.
\end{equation}

We model the force due to lines via a CAK type prescription using the local continuum flux. Working in the Sobolev approximation, the line force
\begin{equation}
\mathbf{G_{\rm{lines}}} = \frac{\rho \kappa_{es}}{c} \sum_{\nu} \oiint M_{\nu}(t) \mathbf{n} I_{\nu}(\mathbf{n}) d\Omega,
\end{equation}
where the integral is over all radiation rays of the continuum in band $\nu$. The frequency dependent force multiplier is a function of the optical depth parameter 
\begin{equation}
    t = \frac{\rho v_{\rm{th}} \sigma_e}{|dv/dl|},
    \label{eq:dvdl}
\end{equation} 
where $v_{\rm{th}} = 4.2 \times 10^{5} \rm{cm/s}$ is the gas thermal velocity and $dv/dl$ the velocity gradient along the line of sight of the radiation flux.

The work done by the line force is then
\begin{equation}
G^0_{\rm{lines}}(E) = v \cdot \mathbf{G_{\rm{lines}}}. 
\end{equation}

We define the Eddington parameter 
\begin{equation}
  \Gamma = L_{*} \sigma_e / 4\pi cGM,
\end{equation}  
where $L_*$ the central object luminosity.

This is the same setup as in DDD25 - we assume gas is optically thin to continuum radiation, and store the local radiation flux for every location in the simulation domain. However, we now store this information for each of the 4 frequency bands i.e the $F_{\nu}$ appearing in equation (\ref{eq:a_nu_final}).

\subsection{Simulation Parameters}
We choose simulation parameters typical for AGN systems. For ease of comparison between models, we will express results as much as possible in terms of dimensionless parameters. We consider black holes in the mass range $3 \times 10^6 \leq M_{\rm{BH}}/M_{\odot} \leq 10^{9}$. The gravitational radius is then $r_g = GM_{\rm{BH}}/c^2$. The inner edge of the disc is assumed to extend to the innermost stable circular orbit, $r_* = 6 r_g$. The inner simulation domain is set to 10$r_*$ to reduce computational limitations of a small time step near the ISCO. Furthermore, previous disc wind simulations have found that wind launching is dominated by radii exterior to this region. We express time in units of the orbital period at the inner radial boundary, $t_0 = 2\pi ((10 r_*)^3/GM)^{1/2}$ with $10 r_*$ the inner radial boundary.

We impose inflow (outflow) boundary conditions at the inner (outer) radial boundaries and axis boundary conditions along the $\theta$ = 0 axis. We assume a reflection symmetry about the $\theta = \pi/2$ midplane.
We use a vacuum boundary condition for the radiation along the disc midplane and outer radial boundaries and keep the ionizing radiation flux fixed at the inner radial boundary. After every full time step we reset $\rho_d = 10^{-8} \rm{g \ cm^{-3}}$, $v_r = 0$ and $v_{\phi} = v_K = \sqrt{GM/r}$. We also impose that the vertical velocity component $v_{\theta}$ is unchanged due to resetting density.

We choose a domain size $n_r$ × $n_{\theta}$ = 96 × 140. The radial domain extends over the range $10 \ r_{*} < r < 500 \ r_{*}$ with geometric spacing $dr_{i+1}/dr_{i} = 1.05$. The polar angle range is $0 < \theta < \pi/2$ and has geometric spacing $d \theta_{j+1}/d \theta{j} = 0.938$, which ensures that we have sufficient resolution near the disc midplane to resolve the acceleration of the flow.

Initially, the cells along the disc are set to have $\rho = \rho_d$, $v_r = v_{\theta} = 0$, $v_{\phi} = v_K$. In the rest of the domain $\rho = 10^{-20} \ \rm{g \ cm^{-3}}$ and $v_r = v_{\theta} = v_{\phi} = 0$. Everywhere, the temperature is constant along vertical cylinders corresponding to the Shakura-Sunyaev disc temperature at the base given by \cite{SS73}.

The disc Eddington fraction $\Gamma_D = 0.3$. We assume the central source does not emit photons that contribute to the line force, but only contribute to the ionization state of the gas. The source of line-driving photons is assumed to extend all the way from the ISCO (effectively outside the simulation domain) with $r_{*} \leq R_d \leq 1500 r_{*}$. A sphere of radius $r_{*}$, effectively the black hole and nearby corona, is assumed to be optically thick to shield the wind from the backside of the disc.  

We impose a density floor $\rho_{\rm{floor}} = 10^{-22} \rm{g \ cm^{-3}}$ which adds matter to stay above this floor, while conserving momentum, if the density ever drops below it. In addition, we have the temperature floor as a function of cylindrical radius.

\section{Results}
\label{sec:results}
We apply our novel method of calculating the radiation force due to spectral lines to study line driven disc winds in AGN. Prior to performing computationally expensive hydrodynamics simulations, we first develop a intuition for how the flux-averaged computation of the line force may affect disc winds models. In Sec. \ref{sec:ForceMultiplierMaps} we produce position dependent maps of the force multiplier for the flux-averaged approximation assuming a wind that is optically thin to the continuum. We compare the radiation force to the method used in DDD25. We then implement this method into the \textsc{Athena++} radiation hydrodynamics code and study the resulting disc wind models which we describe in Sec. \ref{sec:HydroSims}. 

\subsection{Force Multiplier Maps}
\label{sec:ForceMultiplierMaps}

\begin{figure}
    \centering
    \includegraphics[scale=0.66]{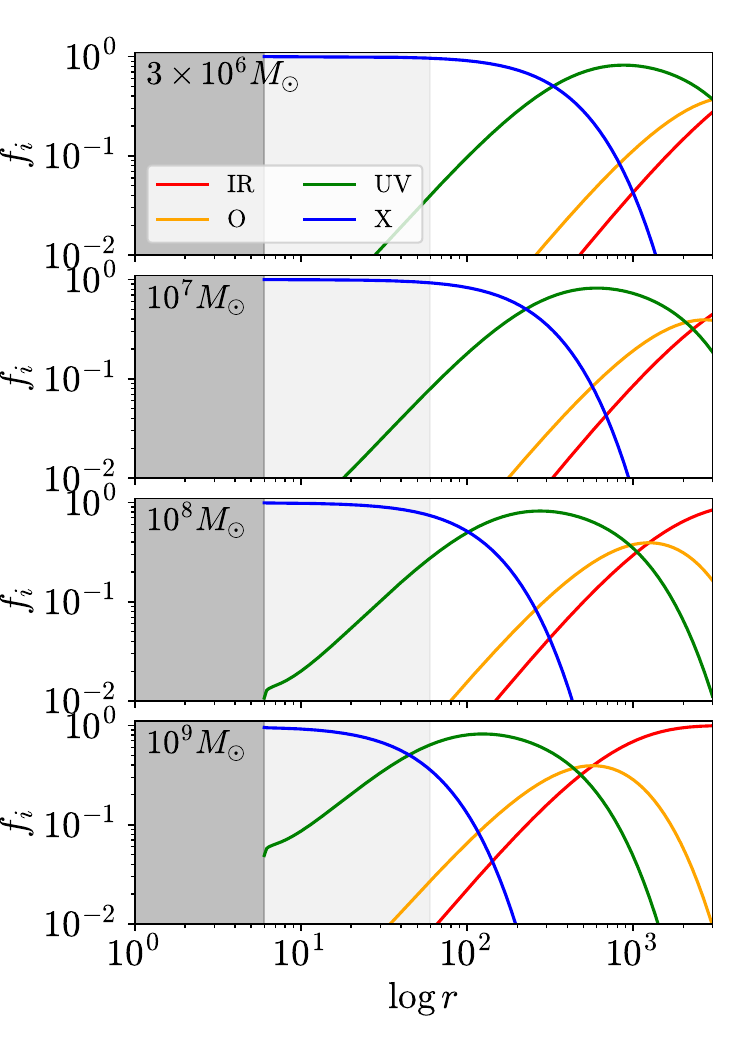}
    \caption{Band fraction as a function of radius along the disc midplane in the IR (red lines), O (orange lines), UV (green lines) and X-ray (blue lines) for black hole masses $M/M_{\odot} = $ $3 \times 10^6$, $10^7$, $10^8$ and $10^9$. The light shaded regions indicate radii interior to the computational domain and the dark shaded region is interior to the ISCO. For higher mass black holes, UV is dominant in the inner parts of the disc whereas for lower masses the X-ray band is dominant.}
    \label{fig:disc_field}
\end{figure}

\begin{figure*}
    \centering
    \includegraphics[scale=0.6]{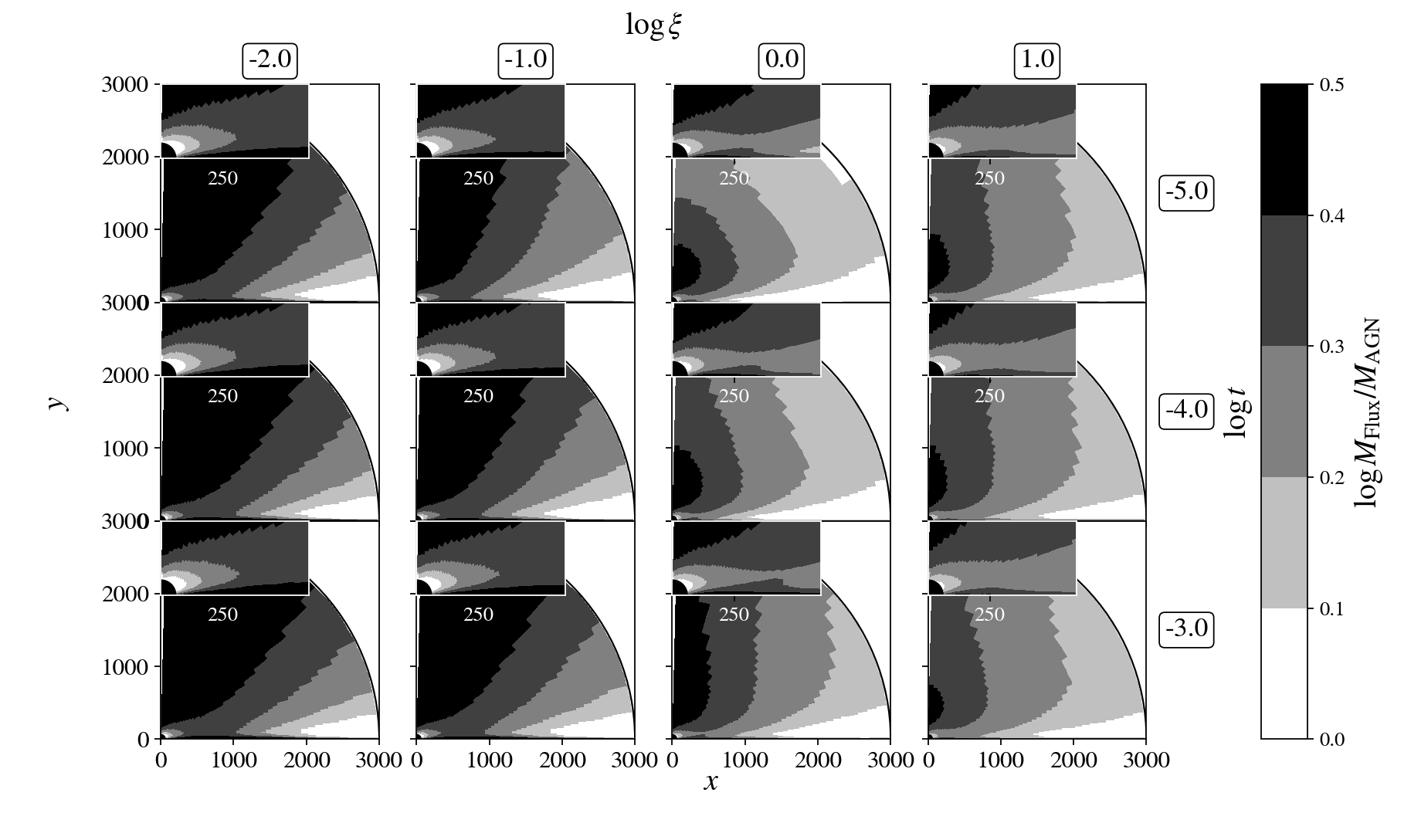}
    \caption{Ratio of the force multiplier $M_{\rm{Flux}}/M_{\rm{AGN}}$ throughout the simulation domain (physical position x-y) for a black hole with mass $M = 10^8 M_{\odot}$. Each column represents a fixed value of the ionization parameter in the range $-2 \leq \log \xi \leq 1$ and each row represents a fixed value of the optical depth parameter in the range $-5 \leq \log t \leq -3$. The flux average force multiplier is enhanced by $\lesssim 2$ in most parts of the domain where wind launching and acceleration occur i.e. near the disc, close to the black hole and in a $45^{\circ}$ wedge above the disc. The enhancement is never more than a factor of 4, and this tends to be in the region above the corona where launching does not occur.}
    \label{fig:FluxAvgDisc}
\end{figure*}

To develop an intuition for the strength of the line force, we compute position dependent maps of the force multiplier for a disc wind geometry. Crucially, we work in the approximation that the wind is optically thin to the continuum and the local radiation flux is thus \emph{time-independent}. We recall from elementary line driven wind theory, \cite{1999isw..book.....L}, that winds can launch when the force multiplier
\begin{equation}
    M(t,\xi) \sim 1/\Gamma.
\end{equation}
For AGN discs, where we expect the disc Eddington fraction to be $\Gamma \sim 10 \%$, the line force is relevant for wind launching when $M(t,\xi) \gtrsim 10$. 

We assume a multi-temperature blackbody disc, where the disc temperature is given by the Shakura-Sunyaev disc temperature. The SED of the disc is highly dependent on the black hole mass. In Fig \ref{fig:disc_field} we plot the fraction of radiation in each band as a function of radius along the disc. Each color represents a black hole with a different mass, with $M/M_{\odot} = $ $3 \times 10^6$ (red lines), $10^7$ (orange lines), $10^8$ (green lines) and $10^9$ (blue lines). Each line style represents a different frequency band - IR (solid lines), O (dash-dot lines), UV (dashed lines) and X-ray (dotted lines). The light shaded regions indicate radii interior to the computational domain and the dark shaded region is interior to the ISCO. For the lowest mass black holes, we see the radiation field is approximately only in the UV and X-rays. In contrast, the highest mass black holes holes have UV, O and IR along the disc and X-rays are emitted almost exclusively from the corona, outside our computational domain.

In the simulation domain we compute the position dependent fluxes in the bands shown in Table \ref{tab:bands}, $F_{\nu}(x,y)$. The flux averaged force multiplier is then
\begin{equation}
    M_{\rm{Flux}} = \sum_{\nu} \frac{F_{\nu}}{F} M_{\nu}(t,\xi). 
\end{equation}
In Fig \ref{fig:FluxAvgDisc} we plot the ratio $M_{\rm{Flux}}/M_{\rm{AGN}}$ throughout the simulation domain (physical position x-y) for a black hole with mass $M = 10^{8} M_{\odot}$ and a variety of optical depth parameters $-5 \leq \log t \leq 5$ and ionizations $-2 \leq \log t \leq 1$. Each panel represents a certain optical depth parameter and ionization parameter of the gas. The contours inside the panel show the ratio of the flux averaged force multiplier and the position independent force multiplier for those gas parameters i.e the ratio of our new treatment of the force multiplier, scaled by the treatment used in DDD25. In other words, the dark grey regions show where we expect a boost in the radiation force. The force multiplier is consistently larger using the flux-averaged prescription though the enhancement is $\lesssim 2$ in most parts of the domain where wind launching and acceleration occur i.e. near the disc, close to the black hole and in a $45^{\circ}$ wedge above the disc. The enhancement is never more than a factor of 4, and this tends to be in the region above the corona where launching does not occur. This is due to geometric effects, whereby the UV photons, which are typically responsible for line driving, dominate only in the innermost parts of the disc. Away from this inner region, where the SED is dominated by longer wavelengths, the line force is weakened. For lower black hole masses, $M \lesssim 10^7 M_{\odot}$, there are less UV photons emitted from the inner parts of the disc, and the enhancement of the force multiplier is somewhat weakened, though still larger than with using the position independent SED. The enhancement for the type 2 AGN is stronger, because the UV fraction in the type 2 SED is less than for the type 1 AGN. Therefore, when computing the flux-average from the disc (which has the same SED in both models), the enhancement which is primarily driven by the UV, receives a larger boost relative to the original SED. In Appendix \ref{sec:FluxAvgMt} we show the equivalent force multiplier contributions broken down by band.

\subsection{Hydrodynamics Simulations}
\label{sec:HydroSims}

\begin{table}
\centering
\begin{tabular}{l c c c c c}
\hline\hline
\multirow{2}{*}{Model} & $\dot{m}$ & $\log \rho_{\rm out}$ & $v_{\rm out}$ & $\omega$ & $\Delta\omega$ \\
& $[M_\odot\,\mathrm{yr^{-1}}]$ & $[\mathrm{g\,cm^{-3}}]$ & $[\mathrm{km\,s^{-1}}]$ & $[^\circ]$ & $[^\circ]$ \\
\hline\hline
Grey   & 0.08 & -16.8 & 5\,500  & 68 & 3  \\ \hline
4 Band & 28.9 & -14.9 & 28\,000 & 54 & 11 \\
O-UV   & 35.7 & -15.2 & 27\,000 & 52 & 7  \\
UV     & 21.5 & -15.5 & 31\,000 & 56 & 8  \\
UV-X   & 18.2 & -15.2 & 32\,000 & 55 & 14 \\ 
\hline\hline
\end{tabular}
\caption{Summary of AGN disc wind models, listing the frequency bands contributing to the line force. We list the time-averaged wind properties for models with an outflow, including the total wind mass flux $\dot{m}$, the typical wind density $\rho_{\rm out}$ and velocity $v_{\rm out}$ at the outer boundary, as well as the wind angle with the zenith $\omega$ and opening angle $\Delta\omega$. The shading indicates the models where $f_{\xi}$ is varied to study the effects of ionization.}
\label{tab:summary}
\end{table}

\begin{figure}
    \centering
    \includegraphics[scale=0.75]{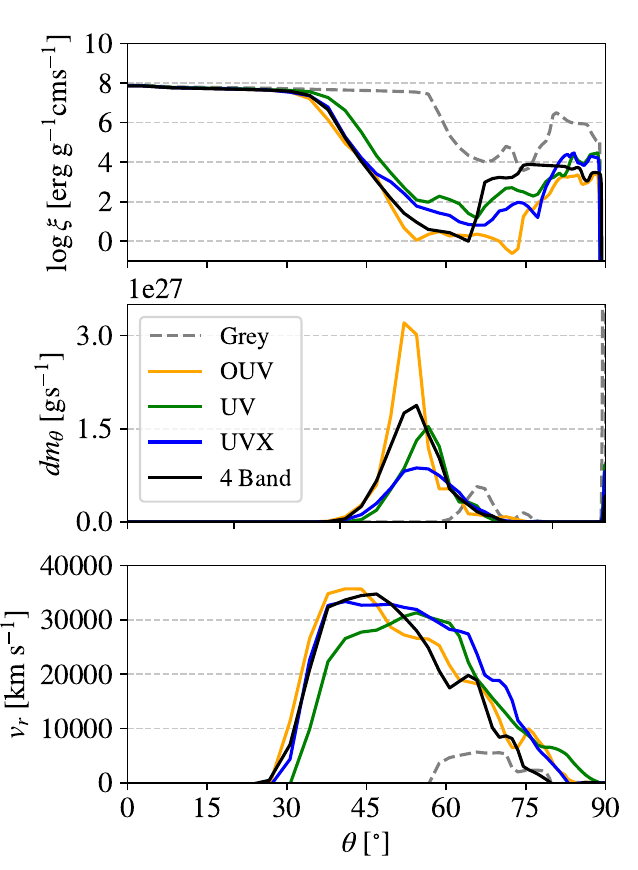}
    \caption{Ionization parameter $\xi$ (top panel), the mass flux density $dm_{\theta} = \rho v_r r^2$ (middle panel) and radial velocity $v_r$ (bottom panel) at the outer boundary for the listed models.  For the mass flux density, the Grey model is shown for x100 for visual clarity.}
    \label{fig:WindOut}
\end{figure}

\begin{figure*}
    \centering
    \includegraphics[scale=0.8]{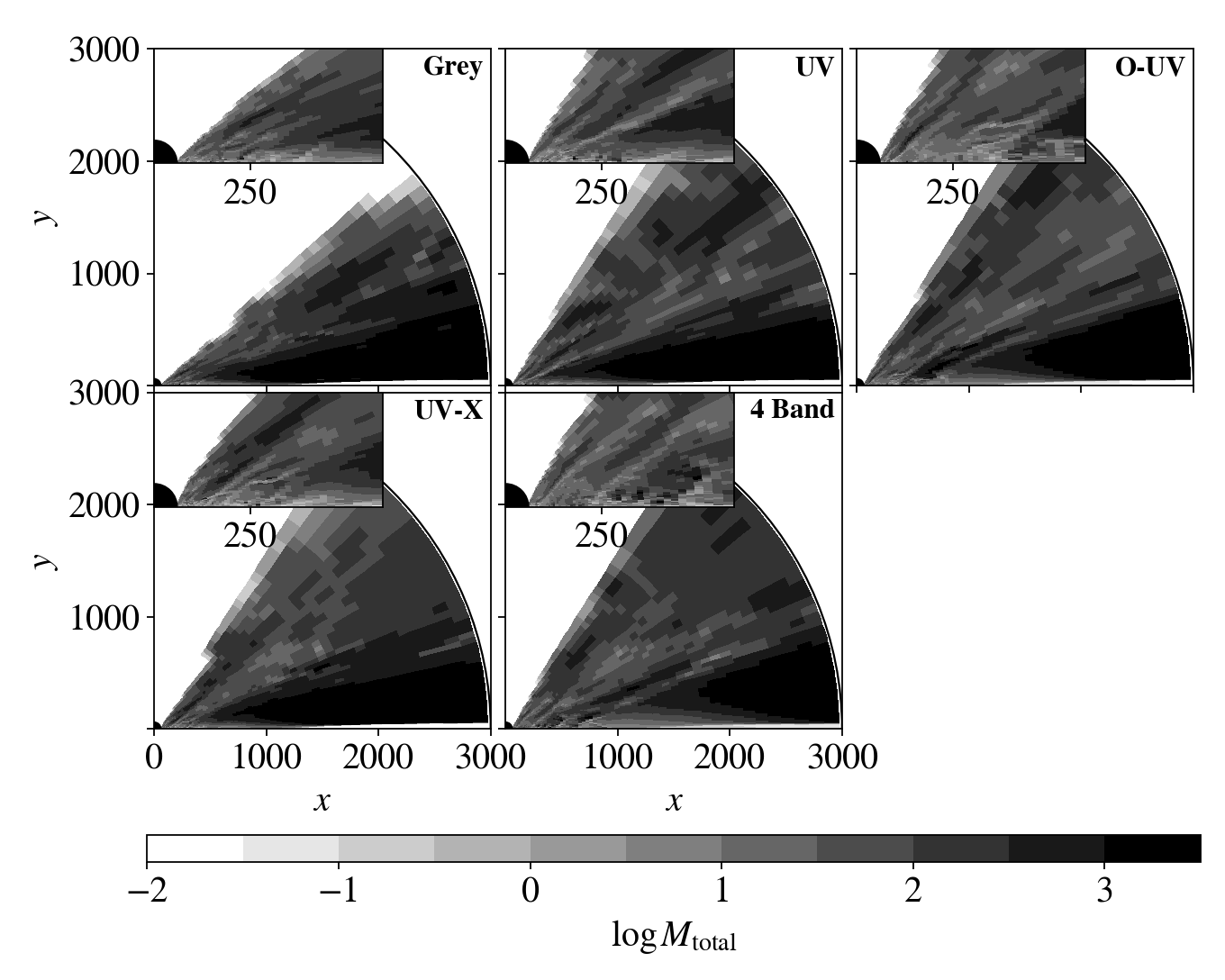}
    \caption{Total force multiplier for models with different bands contributing to the line force.}
    \label{fig:Mt_total}
\end{figure*}

\begin{figure*}
    \centering
    \includegraphics[scale=0.7]{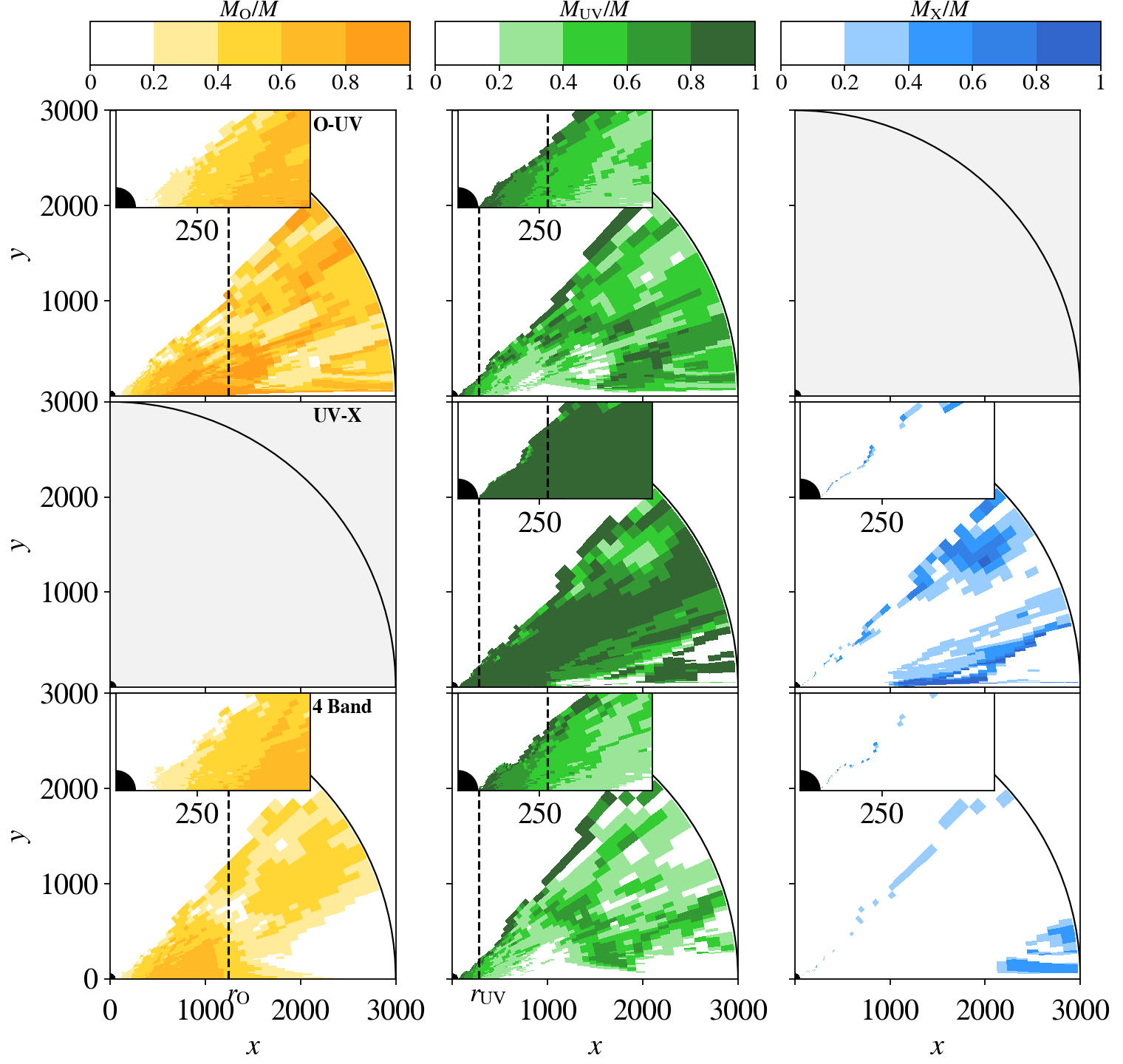}
    \caption{Fraction of force multiplier due to O band (left column), UV (center column), and X-ray (right column) bands for models $O-UV$ (top row), $UV-X$ (middle row) and model 4 Band (bottom row). Using vertical dashed lines we indicate the radius along the disc where emission peaks in the O and UV bands, $r_O$ and $r_{UV}$ respectively. We have omitted the IR band because the force multiplier contribution $M_{IR}/M < 20\%$ throughout the domain.}
    \label{fig:Mt_components}
\end{figure*}

In order to develop an intuition for the effects of each band on the outflow, we conduct a series of simulations where we change which frequency bands contribute to the line force. We have a fiducial model where we use a grey treatment for the line force, that is to say the force multiplier is independent of frequency for all photons contributing to the line force (Grey model). We assume the UV and X fluxes contribute to the line force. This corresponds most closely to the treatment used in DDD25, except for the fact that they used different wavelength cutoffs for the bands that they considered as contributing to the line force (1 \AA $\leq \lambda \leq 3200$ \AA ). We then consider a series of models where we use the band dependent force multipliers. Since the UV band is the dominant contributor to the line force, its force multiplier is always included. In the first model we consider only UV photons. We then consider models with both optical and UV (model O-UV) and both UV and X-ray (model UV-X). Finally, we have a model where we include force multipliers for IR, O, UV and X-ray bands (model 4 Band). This last model`s treatement of the line force corresponds most closely to the fiducial model but with the radiation force now broken down by bands. 

In Fig \ref{fig:WindOut} we plot the ionization parameter $\xi$ (top panel), the mass flux density $dm_{\theta} = \rho v_r r^2$ (middle panel) and radial velocity $v_r$ (bottom panel) at the outer boundary for the listed models. The Grey model has a significantly weaker wind, with velocities a factor a 5 lower than the spectral models and mass fluxes a factor of 100 weaker. This can be understood from the ionization state of the wind, where at the outer boundary the ionization parameter is $\log \xi \gtrsim 5$. At such high ionizations, the force multiplier, irrrespective of band, is barely high enough to launch a wind. The wind morphologies are also different, with the Grey model more radial with a wind zenith angle $\omega = 68^{\circ}$ and narrower opening angle $\Delta \omega \sim 3^{\circ}$.  In contrast, the spectral models have $\omega = 54^{\circ}$ and wider $\Delta \omega \sim 10^{\circ}$. For the Grey model, since the wind is weak, most of the momentum is coming from the innermost parts of the radiation field and the flow is more radial. The spectral models are stronger due to significant radiation forces from the disc, allowing for a vertical component to the force and a raising of the wind angle. They are all morphologically similar, despite slight, factors $\sim 2$ in their outflow velocities and mass fluxes. 

A summary of results from these simulations is listed in Table \ref{tab:summary}. We average over 10 inner disc orbits and list the total mass flux $\dot{m}$ and density $\rho$, velocity $v$, altitude above the midplane $\omega$ and wind opening angle $\Delta \omega$. We measure the density, velocity and altitude at the peak of the mass flux density $dm_{\theta}$ along the outer boundary and define the wind opening angle as the angles where this maximum flux is halved.

In Fig \ref{fig:Mt_total} we plot the total force multiplier as a function of position averaged over 10 inner disc orbits. The Grey model only has strong enough, $M \gtrsim 1/\Gamma$, at the very base of the wind in the inner disc. This leads to a weak, episodic wind, where material is launched during times when the force multiplier becomes strong enough at the base of the wind. In contrast, the spectral models have $\log M \sim 2-3$ at the base of the wind, allowing for robust, sustained wind launching. Further, $\log M \sim 0-1$ further out allowing for material to continue accelerating as the gas escapes the gravitational potential. This allows for considerable, $v \gtrsim 20 \ 000$ km/s winds.  The spectral models benefit from a non-linear feedback effect involving the ionization - a lower ionization parameter increases the force multiplier, launching denser winds and further lowering the ionization parameter.

To understand the effects of different radiation bands we show the fractional contribution to the force multiplier due to each band in Fig \ref{fig:Mt_components}. We show all models with contributions from multiple bands, O-UV (top panels), UV-X (middle panels) and 4 Band (bottom panels). Each column is the respective contribution from the O band (left column), UV (center column) and X (right column). If a band is not included in the model, the panel is left blank. We see that the UV component is dominant in all models near the base of the wind where we know most of the launching occurs. However, we see that the O band is dominant near the disc near radii $r/r_g \sim 1000$. To further understand this, we plot the radii $r_O$ where the blackbody spectrum peaks in the optical band and see that this enhancement in $M_O$ does in fact appear in parts of the wind above $r_O$. Model O-UV does benefit the most from this enhancement and is in fact the strongest wind in terms of mass flux. Crucially, including the O band allows the total force multiplier near $r/r_g \sim 1000$ to be $\log M \sim 3$, whereas for model UV $\log M \sim 2$. 

The contribution of the X-ray band is modest at best. In model UV-X we see that the X band contributes modest $\sim 20\%$ contributions to the force multiplier in the outer parts of the wind. This part of the wind is not responsible for mass loading and only serves to accelerate the wind to greater velocities. We see this in the outflow velocity, where this model has the largest $v \approx 32  \ 000$ km/s. Interestingly, in the 4 Band model, the X band contribution is weak $\lesssim 20 \%$, including in the outer parts of the wind. We note that this is a consequence of the averaging procedure, whereby the X band contribution (as measured relative to UV \emph{and} O) is small. However, looking at the \emph{total} force multiplier in Fig \ref{fig:Mt_total} we note that the 4 Band model has a larger force multiplier than the UV-X model.

\section{Discussion}
\label{sec:discussion}

We have attempted to study line driven disc winds by using the \emph{local, multiband radiation flux} to compute the radiation force. We did this assuming the wind is optically thin to the continuum and that variations in flux do not result in changes to the gas opacity, only to the available momentum in the radiation field that can then be transferred to the gas. This method seeks to leverage the spectral information contained in the photoioinization studies of irradiated gas i.e the \textsc{XSTAR} modeling. The limitation, as previously stated, is that the local SED is used only to compute the momentum transfer but not for computing the gas opacities i.e the force multiplier. 

We note that the models in this paper used different bands compared to our previous study \cite{2025arXiv250400117D}. That work used a \emph{global} force multiplier (the black line in Fig \ref{fig:AGN_4band} for disc photons with 1 \AA $\leq \lambda \leq 3200$ \AA. In contrast, our Grey model used these same tables but with the UV and X bands included, hence 1.24 \AA \ $\leq \lambda \leq $\ 4000 \AA. The closest model to compare to would be their GM1G03x005 with $\dot{m} = 0.15 \rm{M_{\odot}/yr}$ and $v_r = 7 \ 000$ km/s. These are very comparable values ($\dot{m} = 0.08 \rm{M_{\odot}/yr}$ and $v_r = 5 \ 500$ km/s.), so we are confident in these new methods.

The old method of a global SED in a sense artificially capped the contributions of the O band to 4\%, its fraction in the AGN1 SED. By contrast, using the flux-averaged approach, the O band can be dominant in parts of the flow where it is a dominant contributor to the SED i.e near the disc and around $r/r_g \sim 1000$. UV photons are perhaps higher quality for purposes of line driving, in the sense that $M_{UV} > M_O$ (see the distribution of lines Fig \ref{fig:LineDistribution} and the position dependent force multiplier maps \ref{fig:FluxAvgDiscO} and $\ref{fig:FluxAvgDiscUV}$. However, in parts of the flow where $F_O \gg F_{UV}$, the O band may in fact be dominant. Even when they are not dominant, our previous treatment where line driving photons are limited to having a wavelength $\lambda \leq$ 3200\AA \ neglects the possibly crucial role of O band photons. Older models such as PSK00 considered \emph{all} photons as contributing to the line force, while using a global prescription for the force multiplier. The flux-averaged approach offers a way of allowing for O band photons to contribute to the line driving, while also accounting for the fact that different bands couple differently with the gas. 

The ``line mismatch'' problem \citep{Gayley2000} was recognized in the earliest studies of line driven winds in massive stars. It was recognized that for cooler blackbody SEDs, $T_{BB} \lesssim 35 000$ K, there is a mismatch between the peak flux and the frequencies of the strongest lines (see their Fig 4). For AGN SEDs, Fig \ref{fig:LineDistribution} shows that at different ionizations the locations and quality of lines in frequency space can change.  The limitation of the flux-averaged approach is that we assume the coupling is independent of the SED, and future works should seek to improve on this approximation.

One possible approach to attacking the problem of the SED determining both the strength of the gas/radiation coupling (via the line opacity) and the available momentum in the radiation field is to compute \emph{position dependent} force multipliers. In the optically thin approximation, the local radiation field is time-independent, so one could compute a grid of photoionization models for each point in the simulation domain. One would then have a force multiplier ``look-up'' table for every point in the simulation domain. One could potentially reduce the number of \textsc{XSTAR} models needed by resorting to some reasonable approximations about the SED i.e. it is approximately blackbody near the disc and approximately like the AGN spectra ``far'' from the black hole \citep{Smith24} and coupling this to some sort of interpolation scheme across the physical domain. 
This approach would be quite restrictive however, as one would have to recompute all the photoionization models when varying system parameters such as black hole mass. Further, this prescription could not handle any kind of time-dependence from the radiation field, such by say relaxing the optically thin assumption of the wind (to account for scattering) or to model a variable source for example in the context of a reverberation mapping campaign or for changing look AGN. 

An alternative approach is to use SEDs parametrized in the band fractions $f_{\nu}$. One must then compute a larger parameter space of models using their photoionization code. Within the hydrodyanmics simulation, one can read off the local radiation fluxes and extrapolate across these additional parameters when computing the force multiplier. We demonstrated this method using blackbody SEDs for a spherically symmetric wind \citep{2025arXiv250322799D}. Since blackbody SEDs are effectively a family of functions parameterized by a single parameter (the blackbody temperature), this amounted to a modest number of additional photoionization models. However, the built in assumption is that the local SED is approximately blackbody. In our stellar wind problem this approximation is reasonable. However in the AGN problem the radiation field is only blackbody close to the disc. One might then consider two parameter models, say a blackbody and a power-law tail. However, this too is not a great approximation for the SEDs in the AGN wind region. Further, adding additional parameters beyond two will soon make this method computationally infeasible.

Finally, perhaps a machine learning (ML) approach may be applied to this problem. One can imagine computing photoionization models for SEDs parametrized by $\sim$ 10 frequency bands. One can then train a ML model on the photoionization data and this model used to interpolate force multipliers within the hydrodynamics simulation. The key to this approach is to continue adding training data until the required precession is achieved in the force multiplier output. i.e. if we require 5\% deviation from the \textsc{XSTAR} output, then continue adding \textsc{XSTAR} data until this level of precision is achieved. Of course one must take care not to overfit the data and ensure that the level of accuracy is robust.   

The long term goal of this approach is to have an accurate computation of the force multiplier which is \emph{self-consistent} with the local radiation field. This would allow a number of challenging problems to be studied, for example understanding what happens when the optically thin approximation of the wind to the continuum is relaxed (i.e. scattering and re-emission effects), allowing for time-dependent radiation fields (not only in the ionizing photon field) and ultimately time-dependent modeling of the disc, along with a time-dependent radiation field. Ultimately, this type of study is crucial if we are to fully leverage the high resolution, multi-band observations that will become available for AGN from future observatories such as SDSS and the Vera Rubin observatory.

\appendix

\section{Flux Averaged SEDs}
\label{sec:FluxAvgMt}

\begin{figure*}
    \centering
    \includegraphics[scale=0.6]{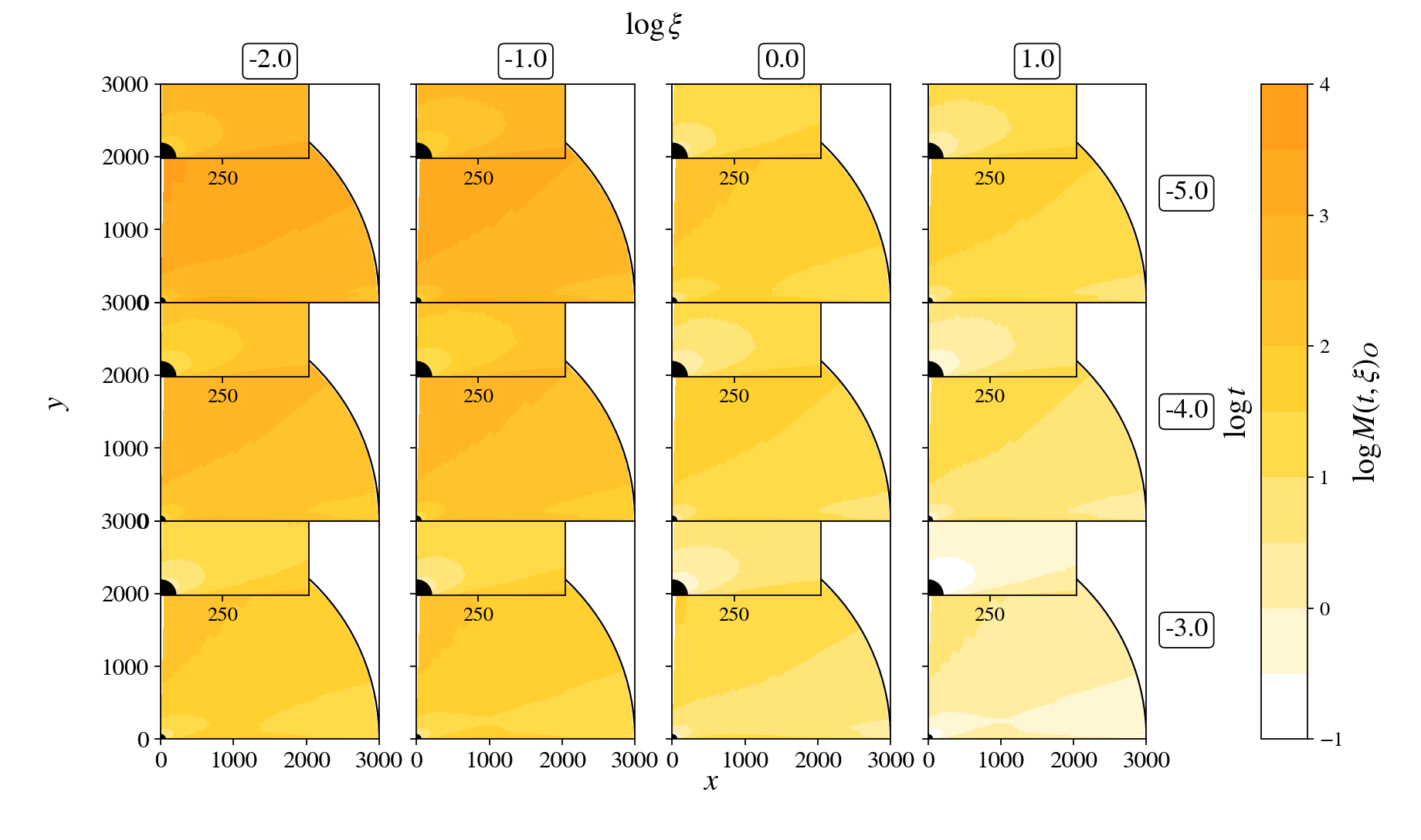}
    \caption{Position dependent force multiplier due to the O band. Each panel represents a different optical depth parameter and ionization parameter of the gas. Here we define the force multiplier as the ratio of the radiation force due to the O band and the radiation force due to electron scattering across the entire SED.}
    \label{fig:FluxAvgDiscO}
\end{figure*}
\begin{figure*}
    \centering
    \includegraphics[scale=0.6]{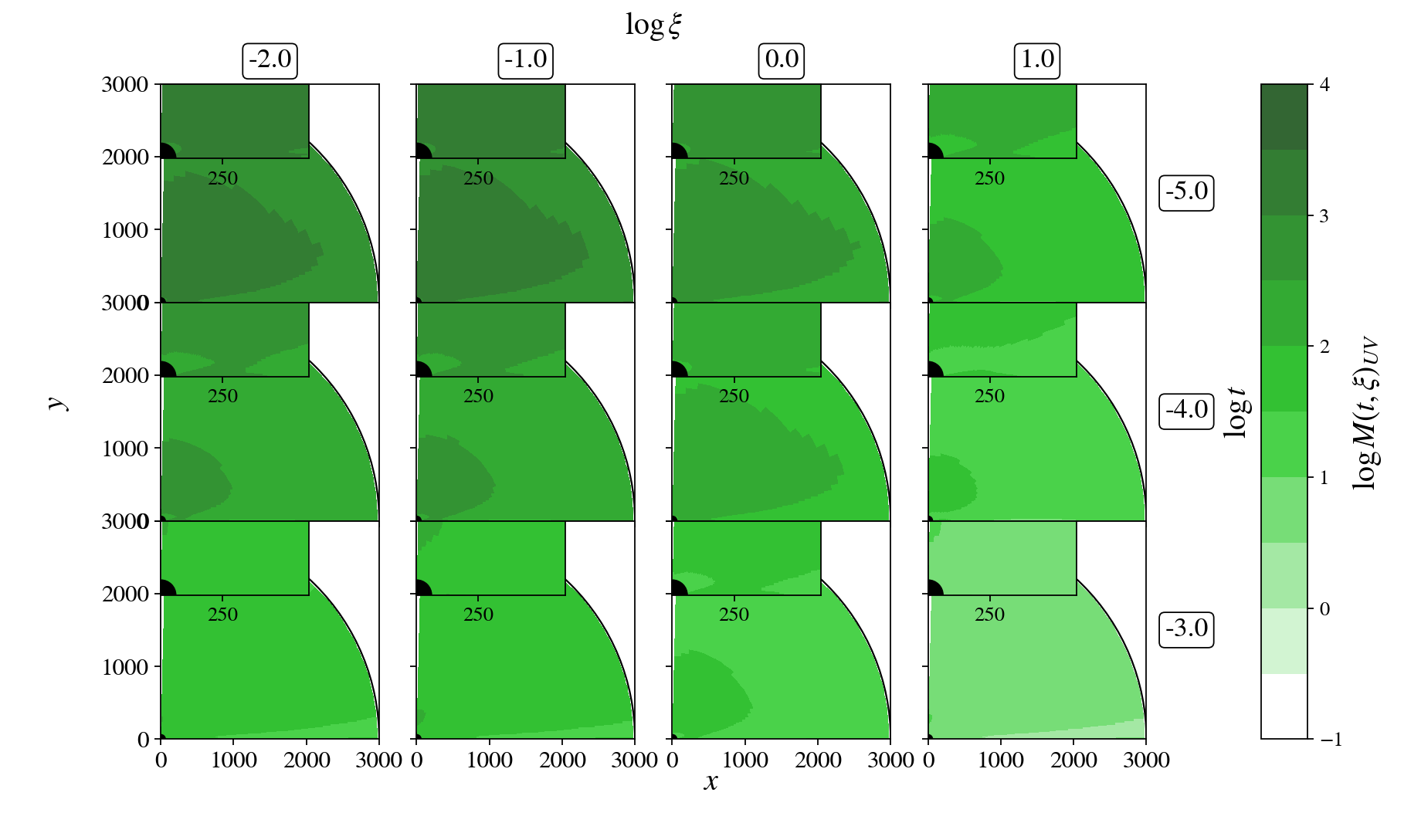}
    \caption{Same as Fig \ref{fig:FluxAvgDiscO} but for the UV band.}
    \label{fig:FluxAvgDiscUV}
\end{figure*}
\begin{figure*}
    \centering
    \includegraphics[scale=0.6]{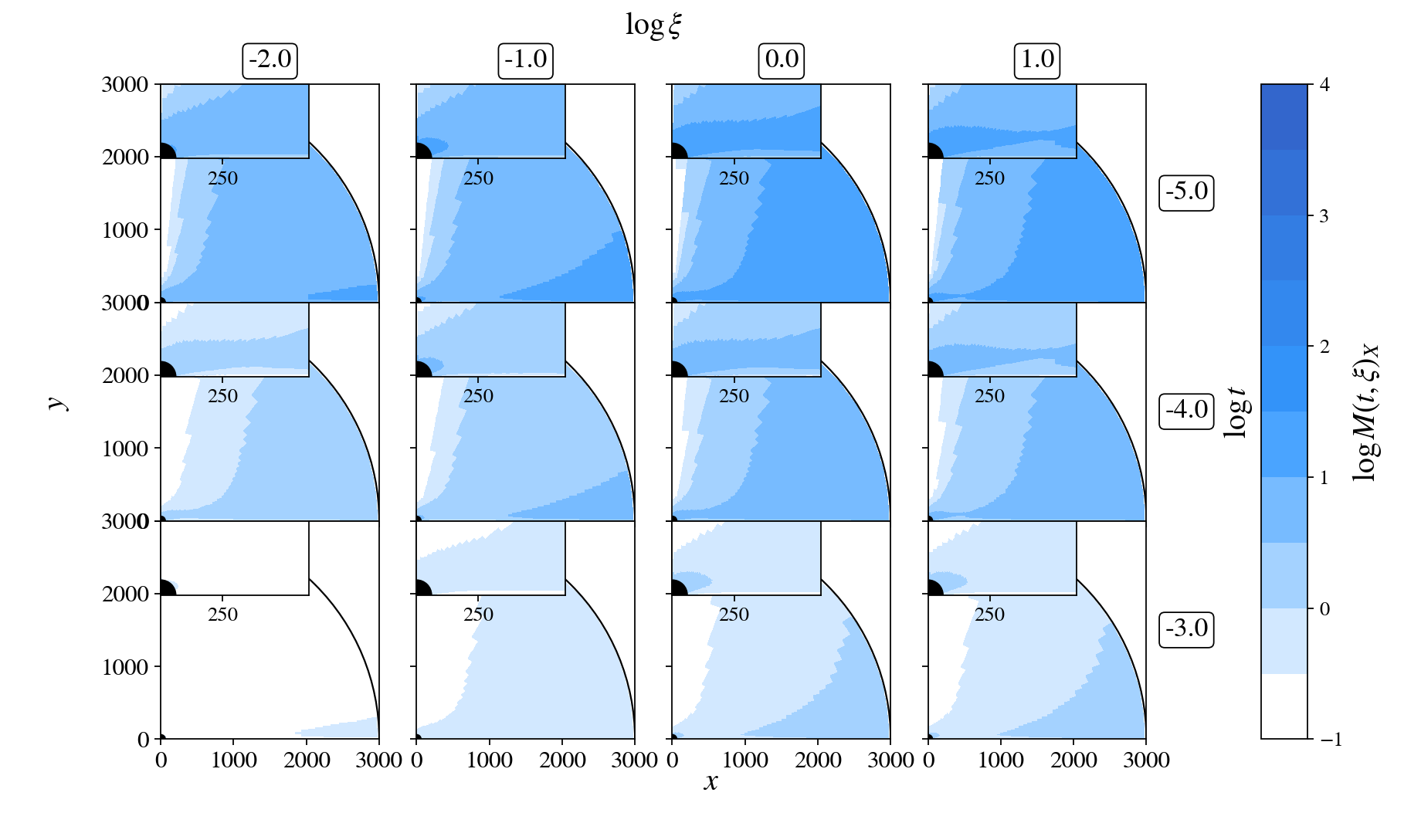}
    \caption{Same as Fig \ref{fig:FluxAvgDiscO} but for the X-ray band.}
    \label{fig:FluxAvgDiscX}
\end{figure*}

We use the frequency dependent force multiplier tables depicted in Fig. \ref{fig:AGN_4band} for unobscursed (AGN1) and obscured (AGN2) AGN to compute position dependent force multipliers. The results are shown for black hole masses $M = 10^{8} M_{\odot}$ for the O band (Fig \ref{fig:FluxAvgDiscO}), UV (Fig \ref{fig:FluxAvgDiscUV}) and X-ray (Fig \ref{fig:FluxAvgDiscX}) bands respectively. Here, we have defined the force multiplier as the ratio of the ratio of the radiation force due to the band of interest, normalized to the radiation force due to electron scattering from the frequency integrated SED. The tables were computed for the IR band as well but the line force is small, $M_{IR} \lesssim 10^{-1}$ in most of the domain for most of the parameter space so we have omitted this plot. 

First we recognize that in most of the parameter space, the line force is dominated by the UV, as we expect from Fig \ref{fig:AGN_4band}. The exception is locations in the domain where the optical flux is larger than the UV flux and simultaneously we are in a part of parameter space where $M_{O} \sim M_{UV}$. Next, we note that the X-ray band is also comparable to the UV driving in certain parts of the parameter space, notably for low optical depth parameter $\log t \sim -6$ and moderate ionizations $\log \xi \sim 2$. This is the contributions of the X-ray lines which were included in the D19 photoionization studies and yielded disc wind enhancements in DDD25. 

For lower mass black holes, the contributions from the X-rays increases. From Fig \ref{fig:AGN_4band}, we see the lower mass black holes are dominated by X-rays in the innermost parts of the disc. Since $I \sim r^{-3}$, this leads to a large enhancement in the contributions of X-rays to the radiation force. In contrast, the higher mass black holes are dominated by the UV contribution. For $M \gtrsim 10^{8} M_{\odot}$, the UV contribution is dominant from $60 \lesssim r/r_g \lesssim 500$, where most of the wind is launched from. Finally, we note that the O band has an important $M_{O} \gtrsim 1$ contribution to line driving in large parts of the parameter space. Including this band is important to accurately compute the line force.



\section*{Acknowledgments} 
Support for this work was provided by the National Aeronautics and Space Administration under TCAN grant 80NSSC21K0496. We thank the entire DAWN TCAN collaboration for fruitful discussions.  The authors acknowledge Research Computing at The University of Virginia for providing computational resources and technical support that have contributed to the results reported within this publication.

\section*{Data Availability Statement}
The simulations were performed with the publicly available code \textsc{Athena++} available at https://github.com/PrincetonUniversity/athena and \textsc{XSTAR} available at https://heasarc.gsfc.nasa.gov/xstar/xstar.html The authors will provide the force multiplier tables to interested parties. The authors will also provide any additional problem generators and input files upon request.


\bibliographystyle{mnras}
\bibliography{progalab-shared}

@article{SS73,
    title          = {{Reprint of 1973A\&A....24..337S. Black holes in binary systems. Observational appearance.}},
    author         = {{Shakura}, N.~I. and {Sunyaev}, R.~A.},
    year           = 1973,
    month          = jun,
    journal        = {\aap},
    volume         = 500,
    pages          = {33--51},
    adsurl         = {https://ui.adsabs.harvard.edu/abs/1973A&A....24..337S}
}

@article{CAK1975,
    title          = {{Radiation-driven winds in Of stars.}},
    author         = {{Castor}, J.~I. and {Abbott}, D.~C. and {Klein}, R.~I.},
    year           = 1975,
    month          = jan,
    journal        = {\apj},
    volume         = 195,
    pages          = {157--174},
    doi            = {10.1086/153315},
    adsurl         = {https://ui.adsabs.harvard.edu/abs/1975ApJ...195..157C}
}

@article{Begelman83,
    title          = {{Compton heated winds and coronae above accretion disks. I. Dynamics.}},
    author         = {{Begelman}, M.~C. and {McKee}, C.~F. and {Shields}, G.~A.},
    year           = 1983,
    month          = aug,
    journal        = {\apj},
    volume         = 271,
    pages          = {70--88},
    doi            = {10.1086/161178},
    adsurl         = {https://ui.adsabs.harvard.edu/abs/1983ApJ...271...70B}
}

@ARTICLE{Blondin1990,
       author = {{Blondin}, John M. and {Kallman}, Timothy R. and {Fryxell}, Bruce A. and {Taam}, Ronald E.},
        title = "{Hydrodynamic Simulations of Stellar Wind Disruption by a Compact X-Ray Source}",
      journal = {\apj},
         year = 1990,
        month = jun,
       volume = {356},
        pages = {591},
          doi = {10.1086/168865},
       adsurl = {https://ui.adsabs.harvard.edu/abs/1990ApJ...356..591B}
}

@article{1991ApJ...379..310S,
    title          = {{X-Ray--illuminated Stellar Winds: Optically Thick Wind Models for Massive X-Ray Binaries}},
    author         = {{Stevens}, Ian R.},
    year           = 1991,
    month          = sep,
    journal        = {\apj},
    volume         = 379,
    pages          = 310,
    doi            = {10.1086/170506},
    adsurl         = {https://ui.adsabs.harvard.edu/abs/1991ApJ...379..310S}
}

@article{1994ApJ...427..700A,
    title          = {{The Role of Radiative Acceleration in Outflows from Broad Absorption Line QSOs. I. Comparison with O Star Winds}},
    author         = {{Arav}, Nahum and {Li}, Zhi-Yun},
    year           = 1994,
    month          = jun,
    journal        = {\apj},
    volume         = 427,
    pages          = 700,
    doi            = {10.1086/174177},
    adsurl         = {https://ui.adsabs.harvard.edu/abs/1994ApJ...427..700A}
}

@article{Blondin1994,
    title          = {{The Shadow Wind in High-Mass X-Ray Binaries}},
    author         = {{Blondin}, John M.},
    year           = 1994,
    month          = nov,
    journal        = {\apj},
    volume         = 435,
    pages          = 756,
    doi            = {10.1086/174853},
    adsurl         = {https://ui.adsabs.harvard.edu/abs/1994ApJ...435..756B}
}

@article{Murray1995,
    title          = {{Accretion Disk Winds from Active Galactic Nuclei}},
    author         = {{Murray}, N. and {Chiang}, J. and {Grossman}, S.~A. and {Voit}, G.~M.},
    year           = 1995,
    month          = oct,
    journal        = {\apj},
    volume         = 451,
    pages          = 498,
    doi            = {10.1086/176238},
    adsurl         = {https://ui.adsabs.harvard.edu/abs/1995ApJ...451..498M}
}

@article{Silk1998,
    title          = {{Quasars and galaxy formation}},
    author         = {{Silk}, Joseph and {Rees}, Martin J.},
    year           = 1998,
    month          = mar,
    journal        = {\aap},
    volume         = 331,
    pages          = {L1-L4},
    archiveprefix  = {arXiv},
    eprint         = {astro-ph/9801013},
    primaryclass   = {astro-ph},
    adsurl         = {https://ui.adsabs.harvard.edu/abs/1998A&A...331L...1S}
}

@book{1999isw..book.....L,
    title          = {{Introduction to Stellar Winds}},
    author         = {{Lamers}, Henny J.~G.~L.~M. and {Cassinelli}, Joseph P.},
    year           = 1999,
    adsurl         = {https://ui.adsabs.harvard.edu/abs/1999isw..book.....L}
}

@ARTICLE{Gayley2000,
       author = {{Gayley}, K.~G. and {Owocki}, S.~P.},
        title = "{Radiative Torque and Partial Spin-Down of Winds from Rotating Hot Stars}",
      journal = {\apj},
         year = 2000,
        month = jul,
       volume = {537},
       number = {1},
        pages = {461-470},
          doi = {10.1086/309002},
       adsurl = {https://ui.adsabs.harvard.edu/abs/2000ApJ...537..461G}
}

@article{PSK2000,
    title          = {{Dynamics of Line-driven Disk Winds in Active Galactic Nuclei}},
    author         = {{Proga}, Daniel and {Stone}, James M. and {Kallman}, Timothy R.},
    year           = 2000,
    month          = nov,
    journal        = {\apj},
    volume         = 543,
    number         = 2,
    pages          = {686--696},
    doi            = {10.1086/317154},
    archiveprefix  = {arXiv},
    eprint         = {astro-ph/0005315},
    primaryclass   = {astro-ph},
    adsurl         = {https://ui.adsabs.harvard.edu/abs/2000ApJ...543..686P}
}

@article{XSTAR2001,
    title          = {{Photoionization and High-Density Gas}},
    author         = {{Kallman}, T. and {Bautista}, M.},
    year           = 2001,
    month          = mar,
    journal        = {\apjs},
    volume         = 133,
    number         = 1,
    pages          = {221--253},
    doi            = {10.1086/319184},
    adsurl         = {https://ui.adsabs.harvard.edu/abs/2001ApJS..133..221K}
}

@article{Chartas02,
    author = {{Chartas}, G. and {Brandt}, W.~N. and {Gallagher}, S.~C. and {Garmire}, G.~P.},
    title          = {{CHANDRA Detects Relativistic Broad Absorption Lines from APM 08279+5255}},
    year           = 2002,
    month          = nov,
    journal        = {\apj},
    volume         = 579,
    number         = 1,
    pages          = {169--175},
    doi            = {10.1086/342744},
    archiveprefix  = {arXiv},
    eprint         = {astro-ph/0207196},
    primaryclass   = {astro-ph},
    adsurl         = {https://ui.adsabs.harvard.edu/abs/2002ApJ...579..169C}
}

@article{2003MNRAS.344..233C,
    title          = {{Continuum shielding and flow dynamics in active galactic nuclei}},
    author         = {{Chelouche}, Doron and {Netzer}, Hagai},
    year           = 2003,
    month          = sep,
    journal        = {\mnras},
    volume         = 344,
    number         = 1,
    pages          = {233--241},
    doi            = {10.1046/j.1365-8711.2003.06841.x},
    archiveprefix  = {arXiv},
    eprint         = {astro-ph/0306513},
    primaryclass   = {astro-ph},
    adsurl         = {https://ui.adsabs.harvard.edu/abs/2003MNRAS.344..233C}
}

@article{PK04,
    title          = {{Dynamics of Line-driven Disk Winds in Active Galactic Nuclei. II. Effects of Disk Radiation}},
    author         = {{Proga}, Daniel and {Kallman}, Timothy R.},
    year           = 2004,
    month          = dec,
    journal        = {\apj},
    volume         = 616,
    number         = 2,
    pages          = {688--695},
    doi            = {10.1086/425117},
    archiveprefix  = {arXiv},
    eprint         = {astro-ph/0408293},
    primaryclass   = {astro-ph},
    adsurl         = {https://ui.adsabs.harvard.edu/abs/2004ApJ...616..688P}
}

@article{Chakravorty09,
    title          = {{Properties of warm absorbers in active galaxies: a systematic stability curve analysis}},
    author         = {{Chakravorty}, Susmita and {Kembhavi}, Ajit K. and {Elvis}, Martin and {Ferland}, Gary},
    year           = 2009,
    month          = feb,
    journal        = {\mnras},
    volume         = 393,
    number         = 1,
    pages          = {83--98},
    doi            = {10.1111/j.1365-2966.2008.14249.x},
    archiveprefix  = {arXiv},
    eprint         = {0811.2404},
    primaryclass   = {astro-ph},
    adsurl         = {https://ui.adsabs.harvard.edu/abs/2009MNRAS.393...83C}
}

@article{2009ApJ...693.1929K,
    title          = {{Three-Dimensional Simulations of Dynamics of Accretion Flows Irradiated by a Quasar}},
    author         = {{Kurosawa}, Ryuichi and {Proga}, Daniel},
    year           = 2009,
    month          = mar,
    journal        = {\apj},
    volume         = 693,
    number         = 2,
    pages          = {1929--1945},
    doi            = {10.1088/0004-637X/693/2/1929},
    archiveprefix  = {arXiv},
    eprint         = {0812.3153},
    primaryclass   = {astro-ph},
    adsurl         = {https://ui.adsabs.harvard.edu/abs/2009ApJ...693.1929K}
}

@article{2010A&A...521A..57T,
    title          = {{Evidence for ultra-fast outflows in radio-quiet AGNs. I. Detection and statistical incidence of Fe K-shell absorption lines}},
    author         = {{Tombesi}, F. and {Cappi}, M. and {Reeves}, J.~N. and {Palumbo}, G.~G.~C. and {Yaqoob}, T. and {Braito}, V. and {Dadina}, M.},
    year           = 2010,
    month          = oct,
    journal        = {\aap},
    volume         = 521,
    pages          = {A57},
    doi            = {10.1051/0004-6361/200913440},
    eid            = {A57},
    archiveprefix  = {arXiv},
    eprint         = {1006.2858},
    primaryclass   = {astro-ph.HE},
    adsurl         = {https://ui.adsabs.harvard.edu/abs/2010A&A...521A..57T}
}

@ARTICLE{Jiang2022,
       author = {{Jiang}, Yan-Fei},
        title = "{Multigroup Radiation Magnetohydrodynamics Based on Discrete Ordinates including Compton Scattering}",
      journal = {\apjs},
         year = 2022,
        month = nov,
       volume = {263},
       number = {1},
          eid = {4},
        pages = {4},
          doi = {10.3847/1538-4365/ac9231},
archivePrefix = {arXiv},
       eprint = {2209.06240},
 primaryClass = {astro-ph.IM},
       adsurl = {https://ui.adsabs.harvard.edu/abs/2022ApJS..263....4J}
}

@article{Fabian2012,
    title          = {{Observational Evidence of Active Galactic Nuclei Feedback}},
    author         = {{Fabian}, A.~C.},
    year           = 2012,
    month          = sep,
    journal        = {\araa},
    volume         = 50,
    pages          = {455--489},
    doi            = {10.1146/annurev-astro-081811-125521},
    archiveprefix  = {arXiv},
    eprint         = {1204.4114},
    primaryclass   = {astro-ph.CO},
    adsurl         = {https://ui.adsabs.harvard.edu/abs/2012ARA&A..50..455F}
}

@article{2013RMxAA..49..137F,
    title          = {{The 2013 Release of Cloudy}},
    author         = {{Ferland}, G.~J. and {Porter}, R.~L. and {van Hoof}, P.~A.~M. and {Williams}, R.~J.~R. and {Abel}, N.~P. and {Lykins}, M.~L. and {Shaw}, G. and {Henney}, W.~J. and {Stancil}, P.~C.},
    year           = 2013,
    month          = apr,
    journal        = {\rmxaa},
    volume         = 49,
    pages          = {137--163},
    archiveprefix  = {arXiv},
    eprint         = {1302.4485},
    primaryclass   = {astro-ph.GA},
    adsurl         = {https://ui.adsabs.harvard.edu/abs/2013RMxAA..49..137F}
}

@article{Pounds13,
    title          = {{The shocked outflow in NGC 4051 - momentum-driven feedback, ultrafast outflows and warm absorbers}},
    author         = {{Pounds}, K.~A. and {King}, A.~R.},
    year           = 2013,
    month          = aug,
    journal        = {\mnras},
    volume         = 433,
    number         = 2,
    pages          = {1369--1377},
    doi            = {10.1093/mnras/stt807},
    archiveprefix  = {arXiv},
    eprint         = {1305.2046},
    primaryclass   = {astro-ph.HE},
    adsurl         = {https://ui.adsabs.harvard.edu/abs/2013MNRAS.433.1369P}
}

@article{Mehdipour15,
    title          = {{Anatomy of the AGN in NGC 5548. I. A global model for the broadband spectral energy distribution}},
    author         = {{Mehdipour}, M. and {Kaastra}, J.~S. and {Kriss}, G.~A. and {Cappi}, M. and {Petrucci}, P. -O. and {Steenbrugge}, K.~C. and {Arav}, N. and {Behar}, E. and {Bianchi}, S. and {Boissay}, R. and {Branduardi-Raymont}, G. and {Costantini}, E. and {Ebrero}, J. and {Di Gesu}, L. and {Harrison}, F.~A. and {Kaspi}, S. and {De Marco}, B. and {Matt}, G. and {Paltani}, S. and {Peterson}, B.~M. and {Ponti}, G. and {Pozo Nu{\~n}ez}, F. and {De Rosa}, A. and {Ursini}, F. and {de Vries}, C.~P. and {Walton}, D.~J. and {Whewell}, M.},
    year           = 2015,
    month          = mar,
    journal        = {\aap},
    volume         = 575,
    pages          = {A22},
    doi            = {10.1051/0004-6361/201425373},
    eid            = {A22},
    archiveprefix  = {arXiv},
    eprint         = {1501.01188},
    primaryclass   = {astro-ph.HE},
    adsurl         = {https://ui.adsabs.harvard.edu/abs/2015A&A...575A..22M}
}

@ARTICLE{Nomura2016,
       author = {{Nomura}, Mariko and {Ohsuga}, Ken and {Takahashi}, Hiroyuki R. and {Wada}, Keiichi and {Yoshida}, Tessei},
        title = "{Radiation hydrodynamic simulations of line-driven disk winds for ultra-fast outflows}",
      journal = {\pasj},
         year = 2016,
        month = feb,
       volume = {68},
       number = {1},
          eid = {16},
        pages = {16},
          doi = {10.1093/pasj/psv124},
archivePrefix = {arXiv},
       eprint = {1511.08815},
 primaryClass = {astro-ph.HE},
       adsurl = {https://ui.adsabs.harvard.edu/abs/2016PASJ...68...16N}
}

@article{Matthews16,
    title          = {{Testing quasar unification: radiative transfer in clumpy winds}},
    author         = {{Matthews}, J.~H. and {Knigge}, C. and {Long}, K.~S. and {Sim}, S.~A. and {Higginbottom}, N. and {Mangham}, S.~W.},
    year           = 2016,
    month          = may,
    journal        = {\mnras},
    volume         = 458,
    number         = 1,
    pages          = {293--305},
    doi            = {10.1093/mnras/stw323},
    archiveprefix  = {arXiv},
    eprint         = {1602.02765},
    primaryclass   = {astro-ph.GA},
    adsurl         = {https://ui.adsabs.harvard.edu/abs/2016MNRAS.458..293M}
}

@article{Dyda17,
    title          = {{Irradiation of astrophysical objects - SED and flux effects on thermally driven winds}},
    author         = {{Dyda}, Sergei and {Dannen}, Randall and {Waters}, Tim and {Proga}, Daniel},
    year           = 2017,
    month          = jun,
    journal        = {\mnras},
    volume         = 467,
    number         = 4,
    pages          = {4161--4173},
    doi            = {10.1093/mnras/stx406},
    archiveprefix  = {arXiv},
    eprint         = {1610.04292},
    primaryclass   = {astro-ph.HE},
    adsurl         = {https://ui.adsabs.harvard.edu/abs/2017MNRAS.467.4161D}
}

@ARTICLE{Matthews2020,
       author = {{Higginbottom}, Nick and {Knigge}, Christian and {Sim}, Stuart A. and {Long}, Knox S. and {Matthews}, James H. and {Hewitt}, Henrietta A. and {Parkinson}, Edward J. and {Mangham}, Sam W.},
        title = "{Thermal and radiation driving can produce observable disc winds in hard-state X-ray binaries}",
      journal = {\mnras},
         year = 2020,
        month = mar,
       volume = {492},
       number = {4},
        pages = {5271-5279},
          doi = {10.1093/mnras/staa209},
archivePrefix = {arXiv},
       eprint = {2001.08547},
 primaryClass = {astro-ph.HE},
       adsurl = {https://ui.adsabs.harvard.edu/abs/2020MNRAS.492.5271H}
}

@ARTICLE{Higginbottom2020,
       author = {{Higginbottom}, Nick and {Knigge}, Christian and {Sim}, Stuart A. and {Long}, Knox S. and {Matthews}, James H. and {Hewitt}, Henrietta A. and {Parkinson}, Edward J. and {Mangham}, Sam W.},
        title = "{Thermal and radiation driving can produce observable disc winds in hard-state X-ray binaries}",
      journal = {\mnras},
         year = 2020,
        month = mar,
       volume = {492},
       number = {4},
        pages = {5271-5279},
          doi = {10.1093/mnras/staa209},
archivePrefix = {arXiv},
       eprint = {2001.08547},
 primaryClass = {astro-ph.HE},
       adsurl = {https://ui.adsabs.harvard.edu/abs/2020MNRAS.492.5271H}
}

@ARTICLE{Nomura2020,
       author = {{Nomura}, Mariko and {Ohsuga}, Ken and {Done}, Chris},
        title = "{Line-driven disc wind in near-Eddington active galactic nuclei: decrease of mass accretion rate due to powerful outflow}",
      journal = {\mnras},
         year = 2020,
        month = may,
       volume = {494},
       number = {3},
        pages = {3616-3626},
          doi = {10.1093/mnras/staa948},
archivePrefix = {arXiv},
       eprint = {1811.01966},
 primaryClass = {astro-ph.HE},
       adsurl = {https://ui.adsabs.harvard.edu/abs/2020MNRAS.494.3616N}
}

@ARTICLE{Harrison2018,
       author = {{Harrison}, C.~M. and {Costa}, T. and {Tadhunter}, C.~N. and {Fl{\"u}tsch}, A. and {Kakkad}, D. and {Perna}, M. and {Vietri}, G.},
        title = "{AGN outflows and feedback twenty years on}",
      journal = {Nature Astronomy},
         year = 2018,
        month = feb,
       volume = {2},
        pages = {198-205},
          doi = {10.1038/s41550-018-0403-6},
archivePrefix = {arXiv},
       eprint = {1802.10306},
 primaryClass = {astro-ph.GA},
       adsurl = {https://ui.adsabs.harvard.edu/abs/2018NatAs...2..198H}
}

@article{Dannen19,
    title          = {{Photoionization Calculations of the Radiation Force Due To Spectral Lines in AGNs}},
    author         = {{Dannen}, Randall C. and {Proga}, Daniel and {Kallman}, Timothy R. and {Waters}, Tim},
    year           = 2019,
    month          = sep,
    journal        = {\apj},
    volume         = 882,
    number         = 2,
    pages          = 99,
    doi            = {10.3847/1538-4357/ab340b},
    eid            = 99,
    archiveprefix  = {arXiv},
    eprint         = {1812.01773},
    primaryclass   = {astro-ph.GA},
    adsurl         = {https://ui.adsabs.harvard.edu/abs/2019ApJ...882...99D}
}

@article{Dannen20,
    title          = {{Clumpy AGN Outflows due to Thermal Instability}},
    author         = {{Dannen}, Randall C. and {Proga}, Daniel and {Waters}, Tim and {Dyda}, Sergei},
    year           = 2020,
    month          = apr,
    journal        = {\apjl},
    volume         = 893,
    number         = 2,
    pages          = {L34},
    doi            = {10.3847/2041-8213/ab87a5},
    eid            = {L34},
    archiveprefix  = {arXiv},
    eprint         = {2001.00133},
    primaryclass   = {astro-ph.GA},
    adsurl         = {https://ui.adsabs.harvard.edu/abs/2020ApJ...893L..34D}
}

@ARTICLE{Matthews2023,
       author = {{Matthews}, James H. and {Strong-Wright}, Jago and {Knigge}, Christian and {Hewett}, Paul and {Temple}, Matthew J. and {Long}, Knox S. and {Rankine}, Amy L. and {Stepney}, Matthew and {Banerji}, Manda and {Richards}, Gordon T.},
        title = "{A disc wind model for blueshifts in quasar broad emission lines}",
      journal = {\mnras},
         year = 2023,
        month = sep,
          doi = {10.1093/mnras/stad2895},
archivePrefix = {arXiv},
       eprint = {2309.14434},
 primaryClass = {astro-ph.GA},
       adsurl = {https://ui.adsabs.harvard.edu/abs/2023MNRAS.tmp.2847M}
}

@ARTICLE{Lucy1970,
       author = {{Lucy}, L.~B. and {Solomon}, P.~M.},
        title = "{Mass Loss by Hot Stars}",
      journal = {\apj},
         year = 1970,
        month = mar,
       volume = {159},
        pages = {879},
          doi = {10.1086/150365},
       adsurl = {https://ui.adsabs.harvard.edu/abs/1970ApJ...159..879L}
}

@ARTICLE{DDP24,
       author = {{Dyda}, Sergei and {Davis}, Shane W. and {Proga}, Daniel},
        title = "{Time-dependent AGN disc winds - I. X-ray irradiation}",
      journal = {\mnras},
         year = 2024,
        month = jun,
       volume = {530},
       number = {4},
        pages = {5143-5154},
          doi = {10.1093/mnras/stae1159},
archivePrefix = {arXiv},
       eprint = {2310.18557},
 primaryClass = {astro-ph.HE},
       adsurl = {https://ui.adsabs.harvard.edu/abs/2024MNRAS.530.5143D}
}

@ARTICLE{2024arXiv240619446R,
       author = {{Robinson}, David and {Avestruz}, Camille and {Gnedin}, Nickolay Y.},
        title = "{On the minimum number of radiation field parameters to specify gas cooling and heating functions}",
      journal = {arXiv e-prints},
         year = 2024,
        month = jun,
          eid = {arXiv:2406.19446},
        pages = {arXiv:2406.19446},
          doi = {10.48550/arXiv.2406.19446},
archivePrefix = {arXiv},
       eprint = {2406.19446},
 primaryClass = {astro-ph.GA},
       adsurl = {https://ui.adsabs.harvard.edu/abs/2024arXiv240619446R}
}

@ARTICLE{Smith24,
       author = {{Smith}, Kara and {Proga}, Daniel and {Dannen}, Randall and {Dyda}, Sergei and {Waters}, Tim},
        title = "{Position-dependent Radiation Fields near Accretion Disks}",
      journal = {\apj},
         year = 2024,
        month = aug,
       volume = {970},
       number = {2},
          eid = {150},
        pages = {150},
          doi = {10.3847/1538-4357/ad4a70},
archivePrefix = {arXiv},
       eprint = {2404.16175},
 primaryClass = {astro-ph.GA},
       adsurl = {https://ui.adsabs.harvard.edu/abs/2024ApJ...970..150S}
}

@ARTICLE{Matthews2025,
       author = {{Matthews}, James H. and {Long}, Knox S. and {Knigge}, Christian and {Sim}, Stuart A. and {Parkinson}, Edward J. and {Higginbottom}, Nick and {Mangham}, Samuel W. and {Scepi}, Nicolas and {Wallis}, Austen and {Hewitt}, Henrietta A. and {Mosallanezhad}, Amin},
        title = "{SIROCCO: a publicly available Monte Carlo ionization and radiative transfer code for astrophysical outflows}",
      journal = {\mnras},
         year = 2025,
        month = jan,
       volume = {536},
       number = {1},
        pages = {879-904},
          doi = {10.1093/mnras/stae2677},
archivePrefix = {arXiv},
       eprint = {2410.19908},
 primaryClass = {astro-ph.HE},
       adsurl = {https://ui.adsabs.harvard.edu/abs/2025MNRAS.536..879M}
}

@ARTICLE{2025arXiv250400117D,
       author = {{Dyda}, Sergei and {Dannen}, Randall C. and {Kallman}, Timothy R. and {Davis}, Shane W. and {Proga}, Daniel},
        title = "{Time-Dependent AGN Disc Winds II -- Effects of Photoionization}",
      journal = {arXiv e-prints},
         year = 2025,
        month = mar,
          eid = {arXiv:2504.00117},
        pages = {arXiv:2504.00117},
          doi = {10.48550/arXiv.2504.00117},
archivePrefix = {arXiv},
       eprint = {2504.00117},
 primaryClass = {astro-ph.HE},
       adsurl = {https://ui.adsabs.harvard.edu/abs/2025arXiv250400117D}
}

@ARTICLE{2025arXiv250322799D,
       author = {{Dyda}, Sergei},
        title = "{Blackbodies Matter}",
      journal = {arXiv e-prints},
         year = 2025,
        month = mar,
          eid = {arXiv:2503.22799},
        pages = {arXiv:2503.22799},
          doi = {10.48550/arXiv.2503.22799},
archivePrefix = {arXiv},
       eprint = {2503.22799},
 primaryClass = {astro-ph.HE},
       adsurl = {https://ui.adsabs.harvard.edu/abs/2025arXiv250322799D}
}

\bsp	
\label{lastpage}

\end{document}